\documentclass[12pt, letterpaper]{article}
\usepackage[toc,page]{appendix}
\usepackage{anysize}
\usepackage{graphics}
\usepackage{color}
\usepackage{amssymb}
\usepackage{graphicx}
\usepackage{epstopdf}
\usepackage[hang,small,bf]{caption}
\usepackage{amsmath}
\usepackage{latexsym}
\usepackage{setspace}
\usepackage{sectsty}
\usepackage{breqn}
\usepackage{tabularx}
\usepackage{comment}

\usepackage{graphicx}
\usepackage{dcolumn}
\usepackage{bm}
\usepackage{epstopdf}

\definecolor{darkred}{cmyk}{0,1,1,0.4}

\usepackage{braket}
\usepackage{xcolor}
\usepackage{slashed}
\usepackage{mathrsfs}
\usepackage{url}
\usepackage[colorlinks,linkcolor=blue,anchorcolor=blue,citecolor=blue, urlcolor=darkred]{hyperref}
\usepackage{caption}

\usepackage{subcaption}
\usepackage{multirow}
\usepackage{float}
\usepackage{amssymb}
\usepackage{enumerate}
\usepackage[marginal]{footmisc}
\usepackage{centernot}
\usepackage{appendix}
\usepackage[compat=1.0.0]{tikz-feynman}
\usepackage[left=2.5cm,right=2cm,top=2.5cm,bottom=2cm]{geometry}
\usepackage{booktabs}
\usepackage{makecell}
\usepackage{scrextend}
\usepackage[normalem]{ulem}
\usepackage{cite}
\usepackage{xspace}

\usepackage{array}
\newcolumntype{L}[1]{>{\raggedright\let\newline\\\arraybackslash\hspace{0pt}}m{#1}}
\newcolumntype{C}[1]{>{\centering\let\newline\\\arraybackslash\hspace{0pt}}m{#1}}
\newcolumntype{R}[1]{>{\raggedleft\let\newline\\\arraybackslash\hspace{0pt}}m{#1}}

\newcommand{\chione}[0]{\tilde{\chi}_1^0}
\newcommand{\chitwo}[0]{\tilde{\chi}_2^0}
\newcommand{\chithree}[0]{\tilde{\chi}_3^0}
\newcommand{\chifour}[0]{\tilde{\chi}_4^0}
\newcommand{\chii}[0]{\tilde{\chi}_i^0}

\newcommand{\chaone}[0]{\tilde{\chi}_1^{\pm}}
\newcommand{\chak}[0]{\tilde{\chi}_k^{\pm}}
\newcommand{\sfermion}[0]{\tilde{f}}
\newcommand{\massplit}[0]{m_{\chitwo} - m_{\chione}}

\newcommand{\slepton}[0]{\tilde{\ell}}
\newcommand{\chargone}[0]{\tilde{\chi}_1^\pm}
\newcommand{\MET}{\ensuremath{\slashed{E}_T}\xspace}
\newcommand{\gmtwo}{\ensuremath{(g_\mu-2)}\xspace}
\newcommand{\ptgamma}{\ensuremath{p_T^\gamma}\xspace}

\newcommand{\eq}[1]{Eq.~(\ref{#1})}
\newcommand{\fig}[1]{Fig.~\ref{#1}}

\newcommand{\sect}[1]{Sec.~\ref{#1}}
\newcommand{\app}[1]{Appendix~\ref{#1}}

\def \beq{\begin{equation}}
\def \eeq{\end{equation}}
\def \bea{\begin{align}}
\def \eea{\end{align}}

\def\lsim{\mathrel{\rlap{\lower4pt\hbox{\hskip1pt$\sim$}}
    \raise1pt\hbox{$<$}}}                
\def\gsim{\mathrel{\rlap{\lower4pt\hbox{\hskip1pt$\sim$}}
    \raise1pt\hbox{$>$}}}                
\renewcommand{\thetable}{\Roman{table}}
\newcolumntype{Y}{>{\centering\arraybackslash}X}

\renewcommand{\thepage}{\arabic{page}}
\renewcommand\thesection{\arabic{section}}

\sectionfont{\large\bf}
\subsectionfont{\normalsize\bf}
\subsubsectionfont{\small\bf}
\begin{document}
\linespread{1}
\title{
\vspace*{-1.5cm}
\begin{flushright}
\normalsize{
FERMILAB-PUB-22-930-T\\
EFI-22-10\\
WSU-HEP-2302}
\end{flushright}
\vspace{0.5cm}
\Large
\textbf{Lighting up the LHC with Dark Matter\\}

\author{\textbf{Sebastian Baum$^{a}$, Marcela Carena$^{b,c,d}$, Tong Ou$^{c}$,} \\
\textbf{Duncan Rocha$^c$, Nausheen R.~Shah$^{e}$, and Carlos E.~M.~Wagner$^{c,d,f}$}\\
[0.3cm]
\normalsize\emph{$^a$ Stanford Institute for Theoretical Physics, Physics Department, Stanford University,} \\ 
\normalsize\emph{Stanford, CA 94305, USA} \\
\normalsize\emph{$^b$~Fermi National Accelerator Laboratory, P.~O.~Box 500, Batavia, IL 60510, USA}\\
\normalsize\emph{$^c$ Enrico Fermi Institute, Physics Department, University of Chicago, Chicago, IL 60637, USA} \\
\normalsize\emph{$^d$ Kavli Institute for Cosmological Physics, University of Chicago, Chicago, IL 60637, USA}\\
\normalsize\emph{$^e$ Department of Physics and Astronomy, Wayne State University, Detroit, MI 48201, USA} \\
\normalsize\emph{$^f$~HEP Division, Argonne National Laboratory, 9700 Cass Ave., Argonne, IL 60439, USA}}
}

\date{}

\maketitle

\begin{abstract}
We show that simultaneously explaining dark matter and the observed value of the muon's magnetic dipole moment may lead to yet unexplored photon signals at the LHC. We consider the Minimal Supersymmetric Standard Model with electroweakino masses in the few-to-several hundred GeV range, and opposite sign of the Bino mass parameter with respect to both the Higgsino and Wino mass parameters. In such region of parameter space, the spin-independent elastic scattering cross section of a Bino-like dark matter candidate in direct detection experiment is suppressed by cancellations between different amplitudes, and the observed dark matter relic density can be realized via Bino-Wino co-annihilation. Moreover, the observed value of the muon's magnetic dipole moment can be explained by Bino and Wino loop contributions. Interestingly, ``radiative'' decays of Wino-like neutralinos into the lightest neutralino and a photon are enhanced, whereas decays into leptons are suppressed. While these decay patterns weaken the reach of multi-lepton searches at the LHC, the radiative decay opens a new window for probing dark matter at the LHC through the exploration of parameter space regions beyond those currently accessible. To complement the current electroweakino searches, we propose searching for a single (soft) photon plus missing transverse energy, accompanied by a hard initial state radiation jet.
\end{abstract}
\newpage

\section{Introduction}
\label{sec:intro}

The origin of the weak scale and the nature of Dark Matter (DM) are arguably the most important and intriguing questions in particle physics. The weak scale is known to be unstable under radiative corrections induced by heavy particles which couple to the Higgs. Many extensions of the Standard Model have been proposed in which these corrections cancel in a natural way. Such extensions often include a discrete symmetry which renders the lightest beyond the Standard Model (BSM) particle stable. These stable particles, which tend to be neutral and weakly interacting, are the prototypical candidates for particle DM and are referred to as WIMPs (Weakly Interacting Massive Particles). 

Low energy Supersymmetry is a very attractive extension of the Standard Model, since it not only solves the stability problem of the electroweak scale and includes a natural DM candidate, but it is also consistent with the unification of gauge couplings at scales close to the Planck scale~\cite{Nilles:1983ge,Haber:1984rc,Martin:1997ns}. The WIMP DM candidate in low energy supersymmetry theories is mostly identified with the lightest {\it neutralino}, the lightest neutral mass eigenstate of the sector comprised of the superpartners of the electroweak gauge and Higgs bosons. Supersymmetric models typically add a multitude of new particles to the SM. However, low energy supersymmetry leads to a rich interplay between different phenomenological aspects. As we will discuss in this work, there are relationships between DM phenomenology, contributions to the anomalous magnetic moments of leptons, and collider phenomenology.

In this work we concentrate on the Minimal Supersymmetric Standard Model (MSSM)~\cite{Nilles:1983ge,Haber:1984rc,Martin:1997ns}. The only option\footnote{Unless one considers non-standard cosmology.} for the MSSM to provide all of the DM with a neutralino in the few-hundred GeV mass range is if that DM candidate is Bino-like, i.e. dominantly composed of the superpartner of the SM's Hypercharge gauge boson. To avoid overclosing the Universe, the annihilation cross section of a Bino-like DM candidate must be enhanced compared to the na\"ive expectation. The classical options discussed in the literature are an appropriate Higgsino admixture (the well-tempered neutralino~\cite{Arkani-Hamed:2006wnf}), resonant $s$-channel annihilation~\cite{Han:2013gba,Cabrera:2016wwr}, annihilation mediated by light staus~\cite{Pierce:2013rda} or smuons~\cite{Fukushima:2014yia} with significant left-right mixing, or co-annihilation with sleptons or charginos~\cite{Ellis:1998kh,Ellis:1999mm,Buckley:2013sca,Han:2013gba,Cabrera:2016wwr,Baker:2018uox,Yanagida:2019evh}.

The most relevant processes for direct detection of a Bino-like DM candidate are controlled by its mixing with the two neutral Higgsinos. The largest contribution to the spin-independent direct detection cross section stems from diagrams mediated by neutral CP-even Higgs bosons; these diagrams are controlled by Bino-Higgsino-Higgs vertices in the interaction basis. The largest contribution to the spin-dependent direct detection cross section stems from $Z$-boson exchange diagrams controlled by Higgsino-Higgsino-$Z$ vertices. During the last decade, the null results from direct detection experiments have put significant pressure on WIMPs in the few-hundred GeV mass region, see, e.g., the current bounds from the XENON1T, PandaX-4T, LZ, and PICO-60 experiments~\cite{XENON:2018voc,PandaX-4T:2021bab,LZ:2022ufs,PICO:2017tgi}. For example, these bounds practically rule out the well-tempered neutralino scenario due to its relatively large Higgsino admixtures to the DM candidate. In order to suppress the direct detection cross section of a Bino-like DM candidate in the MSSM, an obvious possibility is to simply suppress its Higgsino admixtures via a hierarchy between the Higgsino mass parameter $\mu$ and the Bino mass parameter, $|\mu| \gg |M_1|$. However, such a solution is unfavorable from an electroweak finetuning viewpoint since it requires $|\mu|$ to be much larger than the electroweak scale, see, e.g., Ref.~\cite{Drees:2021pbh} for a recent discussion. An alternative albeit much less traveled road is to consider the region of parameter space where $M_1$ and $\mu$ have different signs: for $\left(M_1 \times \mu\right) > 0$, due to the structure of the Higgsino-up and Higgsino-down admixtures with the light Bino, the amplitudes to the spin-independent direct detection cross section mediated by the light and heavy neutral CP-even Higgs bosons add constructively~\cite{Huang:2014xua}. For $\left(M_1 \times \mu\right) < 0$, instead, these amplitudes (partially) cancel~\cite{Huang:2014xua}, leading to a suppression of the spin-independent DM cross section which allows the MSSM to satisfy current direct detection constraints for a Bino-like DM candidate with few-hundred GeV mass while keeping the value of $|\mu|$ comparable to the electroweak scale~\cite{Ellis:2000ds,Ellis:2000jd,Ellis:2005mb,Baer:2006te,Huang:2014xua,Huang:2017kdh,Han:2018gej,Baum:2021qzx}.

As mentioned before, in supersymmetric extensions of the SM, there is rich interplay among different phenomenological characteristics. One aspect that has recently received significant attention are contributions to the anomalous magnetic moment of the muon. With the Fermilab Muon \gmtwo measurement~\cite{Muong-2:2021ojo}, the discrepancy between the measured value and the SM prediction of the muon's anomalous magnetic moment~\cite{Aoyama:2020ynm,Aoyama:2012wk,Aoyama:2019ryr,Czarnecki:2002nt,Gnendiger:2013pva,Davier:2017zfy,Keshavarzi:2018mgv,Colangelo:2018mtw,Hoferichter:2019mqg,Davier:2019can,Keshavarzi:2019abf,Kurz:2014wya,Melnikov:2003xd,Masjuan:2017tvw,Colangelo:2017fiz,Hoferichter:2018kwz,Gerardin:2019vio,Bijnens:2019ghy,Colangelo:2019uex,Blum:2019ugy,Colangelo:2014qya} has reached a statistical significance of $4.2\,\sigma$: $\Delta a_\mu \equiv a_\mu^{\rm exp} - a_\mu^{\rm SM} = \left( 25.1 \pm 5.9 \right) \times 10^{-10}$, where $a_\mu \equiv (g_\mu - 2)/2$. We note that this discrepancy is based on the theoretical estimate of $a_\mu^{\rm SM}$ as published in Ref.~\cite{Aoyama:2020ynm}; in particular, this estimate uses an extraction of the hadronic vacuum polarization (HVP) contribution to $a_\mu^{\rm SM}$ from ($e^+ e^- \to {\rm hadrons}$) cross section measurements via dispersion relations. Currently, there is a disagreement between this data-driven determination of the HVP contribution and {\it ab initio} lattice calculations of the HVP contribution~\cite{Borsanyi:2020mff}. Despite significant effort, see, e.g., Refs.~\cite{SND:2020nwa,Aubin:2022hgm,Colangelo:2022vok,Ce:2022kxy,Alexandrou:2022amy,Colangelo:2022lzg,FermilabLattice:2022izv,Alexandrou:2022tyn,Gagliardi:2022bxo,Bazavov:2023has,Blum:2023qou,CMD-3:2023alj}, this discrepancy is not resolved at the time of writing this article; in the following we will assume that the data-driven determination of the HVP contribution is correct. 

It is well known that the MSSM can provide contributions to $a_\mu$ which account for the difference between the SM prediction and the measured value~\cite{Barbieri:1982aj,Ellis:1982by,Kosower:1983yw,Moroi:1995yh,Carena:1996qa,Feng:2001tr,Martin:2001st,Marchetti:2008hw,Athron:2015rva,Endo:2019bcj,Badziak:2019gaf,Abdughani:2019wai,Chakraborti:2020vjp,Yin:2021mls,Chakraborti:2021kkr,Chakraborti:2021dli,Baer:2021aax,Chakraborti:2021mbr,Baum:2021qzx}. The most important contributions arise from a Bino-smuon and a chargino-(muon-sneutrino) loop. Interestingly, the sign of these contributions to $a_\mu$ is controlled by the signs of $\left(M_1 \times \mu \right)$ and $\left(M_2 \times \mu\right)$, respectively. For $\left(M_1 \times \mu\right) < 0$, motivated by realizing a viable DM candidate without exacerbating electroweak finetuning issues, the Bino-smuon contribution to $a_\mu$ is negative, while the measured value $a_\mu^{\rm exp}$ is larger than its SM prediction. Nonetheless, the parameters of the MSSM can easily be arranged to provide an overall positive contribution to $a_\mu$ as long as $\left(M_2 \times \mu \right)$ is positive~\cite{Baum:2021qzx}. In order for the MSSM contributions to explain the measured value of $a_\mu$, the Binos, Winos, and smuons must not be too heavy. For example, for a value of the ratio of the vacuum expectation values of the down-type and the up-type Higgs bosons of $\tan\beta = v_u/v_d = 10$, one obtains the observed value of $a_\mu$ if Binos, Winos, and smuons all have masses of order $\widetilde{m} \sim 200\,$GeV, while for $\tan\beta = 60$, one obtains the observed $a_\mu$ if $\widetilde{m} \sim 500\,$GeV. Thus, requiring the MSSM to provide not only a viable DM candidate but also to explain the \gmtwo anomaly further motivates considering a region of the parameter space where Binos, Winos, and sleptons have masses of a few hundred~GeV.

Summarizing the discussion in the preceding paragraphs, realizing a DM candidate that satisfies current direct detection bounds without exacerbating the electroweak finetuning problem and simultaneously explaining the observed value of the muon's magnetic dipole moment prefers a region of the MSSM parameter space where the Bino mass parameter $M_1$, the Wino mass parameter $M_2$, and the Higgsino mass parameter $\mu$ all take (absolute) values of a few hundred GeV, and where their signs take the particular combination $(M_1 \times \mu) < 0$ and $(M_2 \times \mu) > 0$, implying $(M_1 \times M_2) < 0$. The most straightforward option to explain the observed DM relic density is then co-annihilation of a Bino-like DM candidate with the Wino-like neutralinos and charginos, corresponding to the parameter region where $M_1$ is smaller (in magnitude) than $M_2$, but where $\left|M_2\right|$ is not much larger than $\left|M_1\right|$. As we will see, a mass splitting between the Bino-like lightest neutralino ($\chione$) and a Wino-like second lightest neutralino ($\chitwo$) of $(\massplit) \sim 10-30\,$GeV leads to our DM candidate $\chione$ explaining the observed relic density of our Universe. In the remainder of this article, we will refer to the region of parameter space where the Bino-Wino mass splitting is in this range as the {\it compressed region}. 

\begin{figure}
    \centering
    \includegraphics[width=0.5\linewidth]{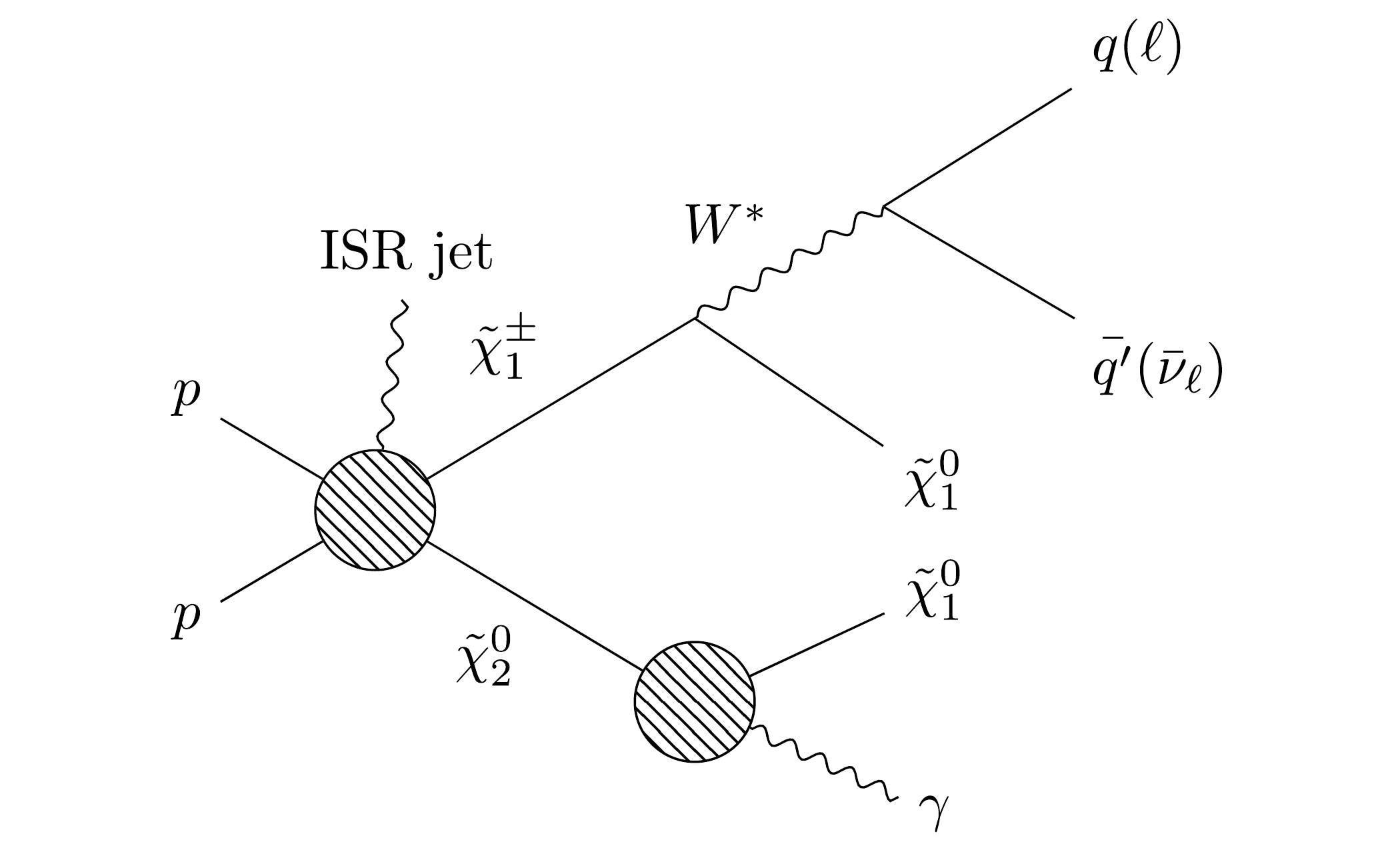}
    \caption{Illustration of a representative process giving rise to the mono-photon~+~\MET~+~jets/leptons final state arising at the LHC via radiative decays of the Wino-like neutralino $\chitwo$.}
    \label{fig:ChannelFeyn}
\end{figure}

The Large Hadron Collider (LHC) is the premier tool to directly search for new particles. To date, searches for new particles at the LHC have yielded null results, setting relevant constraints on the MSSM parameter space, in particular, requiring new color-charged particles such as gluinos ($\widetilde{g}$) and squarks ($\widetilde{q}$) to have masses $m_{\widetilde{g}} \gtrsim 2\,$TeV and $m_{\widetilde{q}} \gtrsim 1\,$TeV, respectively, see, e.g., Refs.~\cite{ATLAS:2017tmw,ATLAS:2018nud,CMS:2019zmd,CMS:2019ybf,ATLAS:2020syg,ATLAS:2021twp,ATLAS:2021kxv}.\footnote{We note that this mass region is also preferred by the observed 125\,GeV mass of the SM-like Higgs boson. Radiative corrections dominated by stops are required to lift the mass of a SM-like Higgs to such values in the MSSM; reproducing $m_h \sim 125\,$GeV requires stops with at least few-TeV masses.} With the analyses of the LHC Run~2 data, the ATLAS and CMS collaborations have also started to provide interesting bounds on new color-neutral particles such as Binos, Winos, and sleptons in the few-hundred GeV mass range. The progress in searches using soft multi-lepton + missing transverse energy (\MET) final states aimed at the compressed region has been particularly impressive~\cite{ATLAS:2018ojr,ATLAS:2019lff,CMS:2020bfa,ATLAS:2021moa,CMS:2021cox,CMS:2021few,ATLAS:2021yqv,ATLAS:2022zwa,CMS:2022sfi}. Such searches are well motivated: in a DM-motivated scenario where Binos and Winos are relatively light and $\left|M_1\right| < \left|M_2\right|$, Wino pair production cross sections at the LHC are sizeable, especially in the ($pp \to \chitwo~\chaone$) channel where $\chitwo$ and $\chaone$ are the Wino-like neutralino and chargino, respectively, and the branching ratios of $(\chitwo \to \chione + f \bar{f})$ processes are typically quite large. Alternative searches targeting the compressed region using the vector boson fusion signatures can provide additional probes and comparable constraints, although the current constraints are weaker than those considered in this work, see e.g. Ref.~\cite{CMS:2019san}, which also assumes leptonic decays of the Wino-like neutralino and chargino. However, as we discuss in this work, there is an interesting interplay between the sign of $\left(M_1 \times M_2 \right)$ and the decay modes of the Wino-like neutralinos: for $\left(M_1 \times M_2\right) > 0$, radiative decays ($\chitwo \to \chione + \gamma$) mediated by loops involving charginos or sleptons have relatively small branching ratios and the $(\chitwo \to \chione + f \bar{f})$ decays dominate~(except in the very compressed regime where $m_{\chitwo} \simeq m_{\chione}$). Instead for $\left( M_1 \times M_2 \right) < 0$, the different diagrams mediating ($\chitwo \to \chione + \gamma$) decays interfere constructively, enhancing the associated branching ratio and suppressing the $(\chitwo \to \chione + f \bar{f})$ decays.

Recall that realizing a DM candidate in the MSSM compatible with direction detection constraints for moderate values of $|\mu|$ prefers $(M_1 \times \mu) < 0$, and providing a correction to the muon's anomalous magnetic moment that explains the discrepancy between the SM prediction and the measured value requires $(M_2 \times \mu) > 0$. Hence, $(M_1 \times M_2) < 0$, which leads to large radiative branching ratios of the second-lightest neutralino ($\chitwo \to \chione +\gamma$); we quite generally find values of order ${\rm BR}(\chitwo \to \chione +\gamma) \sim 0.2 - 0.4$. On the one hand, this means that the existing multi-lepton+\MET searches are less sensitive in this region. On the other hand, the large radiative-decay branching ratios motivate a new search channel: mono-photon+\MET accompanied by jets or a lepton from the chargino decay and possible initial state radiation (ISR), see \fig{fig:ChannelFeyn}. Interestingly, during the long shutdown of the LHC preceding the current LHC Run~3, the experimental collaborations have significantly upgraded the event selection triggers.\footnote{We thank David Miller for a discussion on these issues.} These upgrades now allow one to trigger on a combination of a single photon and \MET with much lower transverse-momentum ($p_T$) thresholds than a more traditional photon-only or \MET-only trigger would allow for. In combination with expected improvements in the multi-lepton+\MET searches with LHC Run~3 data, this opens up the exciting possibility to make significant progress towards probing the region of the MSSM parameter space motivated by DM and the muon \gmtwo anomaly in the near future, and perhaps, to make a discovery.

We note that some of the qualitative behavior of the region of parameter space we consider here has been discussed in Ref.~\cite{Baer:2005jq}. In particular, that work highlights the interplay between the suppression of the direct detection cross section, the Bino-Wino co-annihilation mechanism, and the enhancement of the radiative decay branching ratio of the second-lightest neutralino in the compressed Bino-Wino region if $(M_1 \times M_2) < 0$ and $(M_1 \times \mu) < 0$. Since Ref.~\cite{Baer:2005jq} was written years priors to the start of LHC operations, that work considered much smaller masses of the strongly charged superpartners than what is allowed by current LHC searches for squarks and gluinos as well as what is required by realizing a 125\,GeV SM-like Higgs boson. For the squark and gluino masses considered in Ref.~\cite{Baer:2005jq}, the dominant production mode of Wino-like neutralinos at the LHC would have been in the decay cascades of gluinos and squarks, hence, Ref.~\cite{Baer:2005jq} proposed to search for events containing isolated photons from the radiative decays of the Wino-like neutralinos and the strongly charged SM decay products of the gluinos and squarks. In the region of parameter space we consider in this work, squarks and gluinos are much heavier and hence their production cross sections much smaller. For example, for $\sim 2.5\,$TeV squark and gluino masses as we will assume later on, the gluino pair production cross section at the $\sqrt{s} = 13\,$TeV LHC is $\sim 0.08\,$fb, the squark+antisquark production cross section (assuming 10 degenerate squark species) is $\sim 0.01\,$fb, and the gluon+squark and squark+squark production cross sections are $\sim 0.5\,$fb~\cite{Borschensky:2014cia}, while for the squark and gluino masses of $\lesssim 750\,$GeV considered in Ref.~\cite{Baer:2005jq}, each of these production cross sections would be larger than 10\,pb. For the scenario considered in this work, the dominant production mode of Wino-like charginos and neutralinos at the LHC is direct Wino pair production, then, radiative decays of the second-lightest neutralino give rise to the mono-photon + \MET signature we propose as a new LHC search channel.

This article is organized as follows. In \sect{sec:spec} we describe the neutralino, chargino and slepton sectors relevant for the present analysis. In \sect{sec:analytical} we discuss the phenomenological properties of the MSSM in the compressed region, presenting analytical results that contribute to the understanding of the numerical results of our work. In \sect{sec:Computations} we present our numerical results, and in \sect{sec:LHC} we discuss their relevance for current and future searches for electroweakinos at the LHC. We reserve \sect{sec:conclusions} for our conclusions. Additional plots and some analytical formulae are presented in the Appendices.

\section{Neutralino, Chargino, and Slepton Sectors of the MSSM}
\label{sec:spec}

In this section, we introduce the properties of the electroweak sector of the MSSM, which will be the focus of the study performed in this article. In particular, we present the most relevant parameters defining the electroweakino and slepton masses and mixing angles, and define notations. Some of these parameters were already presented in the introduction, but we reiterate their meanings here to make this a self-contained section.

The neutralino sector of the MSSM consists of the neutral superpartners of the electroweak $SU(2) \times U(1)_Y$ gauge bosons and of the Higgs doublets. Similarly, the chargino sector is composed of the superpartners of the charged Higgs and gauge bosons. Due to gauge invariance, there are two independent supersymmetry-breaking mass parameters which we denote $M_1$ and $M_2$, giving masses to the Hypercharge gaugino (Bino) and to the neutral and charged $SU(2)$ Winos. An additional mass parameter, $\mu$, appears in the supersymmetric Lagrangian, giving masses to the neutral and charged Higgsinos. After electroweak symmetry breaking, the gaugino and Higgsino states mix through their interactions with the Higgs fields; the mixing term is controlled by the electroweak gauge couplings and the two Higgs vacuum expectation values. The mass matrix for the (Majorana) neutralino states is given by
\begin{align}
    \mathbf{M}_{N}   =&  \begin{pmatrix}
        M_{1} & 0 & -c_{\beta}s_{W}m_{Z} & s_{\beta}s_{W}m_{Z}\\
        0 & M_{2} &c_{\beta}c_{W}m_{Z} & -s_{\beta}c_{W}m_{Z}\\
        -c_{\beta}s_{W}m_{Z} & c_{\beta}c_{W}m_{Z} & 0 & -\mu\\
        s_{\beta}s_{W}m_{Z} & - s_{\beta}c_{W}m_{Z} & -\mu & 0
        \end{pmatrix}.
        \label{eq:mn_matrix}
\end{align}
Here, $\tan\beta \equiv v_u/v_d$ is the ratio of the two Higgs vacuum expectation values associated with the Higgs bosons which lead to the up-type and down-type quark masses at tree-level, $H_u$ and $H_d$. We use the shorthand notation $c_\beta \equiv \cos\beta$, $s_\beta \equiv \sin\beta$; $m_{Z}$ and $m_{W}$ are the $Z$ and $W$ gauge boson masses; $\theta_{W}$ is the weak-mixing angle, and $c_{W} \equiv \cos\theta_{W}$, $s_{W} \equiv \sin\theta_{W}$. The mass matrix $\mathbf{M}_N$ can be diagonalized by an orthogonal matrix $\mathbf{N}$ to obtain mass eigenstates and eigenvalues:
\begin{equation}
    \chii=\mathbf{N}_{ij}\psi_j^0 \;,
\end{equation}
\begin{equation}
\label{eq:2.4}
    \mathbf{N}^{\ast}\mathbf{M}_{N}\mathbf{N}^{-1}={\rm diag}(\xi_1 m_{\chione},\xi_2 m_{\chitwo},\xi_3 m_{\chithree},\xi_4 m_{\chifour}) \;,
\end{equation}
where $\psi_j^0=(\widetilde{B},\widetilde{W}^0,\widetilde{H}_d^0,\widetilde{H}_u^0)$ is the gauge-eigenstate basis, and the eigenvalues ($\xi_i m_\chii$) can be positive or negative. In Eq.~(\ref{eq:2.4}) we define $m_\chii$ to be positive and absorb the sign obtained from the diagonalization into $\xi_i$. 

The mass matrix for the chargino states is given by
\begin{equation}
     \mathbf{M}_{C} =\begin{pmatrix}
        \mathbf{0} & \mathbf{X}^T\\
        \mathbf{X} & \mathbf{0}
        \end{pmatrix} \,,
\end{equation}
with
\begin{equation}
    \mathbf{X}= \begin{pmatrix}
        M_{2} & \sqrt{2}s_{\beta}m_{W}\\
        \sqrt{2}c_{\beta}m_{W} & \mu
        \end{pmatrix} \,.
\end{equation}
The mass eigenstates are related to the gauge-eigenstate basis via
\begin{equation}
    \begin{pmatrix}
        \tilde{\chi}_1^+ \\ \tilde{\chi}_2^+
    \end{pmatrix}
    =\mathbf{V}\begin{pmatrix}
        \widetilde{W}^+ \\ \widetilde{H}_u^+
    \end{pmatrix},\qquad \begin{pmatrix}
        \tilde{\chi}_1^- \\ \tilde{\chi}_2^-
    \end{pmatrix}
    =\mathbf{U}\begin{pmatrix}
        \widetilde{W}^- \\ \widetilde{H}_d^-
    \end{pmatrix},
\end{equation}
where $\mathbf{V}$ and $\mathbf{U}$ satisfy
\begin{equation}
    \mathbf{U}^{\ast} \mathbf{X}\mathbf{V}^{-1}={\rm diag}(m_{\tilde{\chi}_1^{\pm}},m_{\tilde{\chi}_2^{\pm}}) \;.
\end{equation} 
For further details of the couplings and mass matrices of neutralinos and charginos see, for instance, Ref.~\cite{Martin:1997ns}.

The slepton sector, on the other hand, consists of the superpartners of the left- and right-handed lepton fields. The diagonal entries to the slepton mass matrix (in the chirality basis) are controlled by the supersymmetry-breaking terms $M_L^2$ and $M_R^2$, associated with the left- and right-handed sleptons, respectively, and the so-called $D$-terms which are proportional to the square of the gauge couplings and the Higgs vacuum expectation values. The mixing of the left- and right-handed sleptons receives contributions from two sources: first, the supersymmetry-breaking $A_l$-terms between left-handed sleptons, right-handed sleptons, and $H_d$. Assuming minimal flavor violation, we will use a universal $A_l$ for all generations and the resulting mixing term is proportional to the SM Yukawa coupling of the lepton associated with the slepton. Second, the supersymmetric Lagrangrian introduces another trilinear term between left-handed sleptons, right-handed sleptons and $H_u$; this trilinear term is proportional to the Higgsino mass parameter $\mu$. Assuming no inter-generational mixing, the mass matrix for a given slepton generation is given by
\begin{align} \label{eq:Mslep}
    \mathbf{M}_{\tilde{l}}^2
    = \begin{pmatrix} 
        M_L^2 +m_l^2 + D_L & m_l(A_l -\mu \tan\beta) \\
        m_l (A_l -\mu\tan\beta) & M_R^2 + m_l^2 + D_R
    \end{pmatrix},
\end{align}
where $m_l$ is the SM lepton mass, $D_L = (-1/2+s^2_W) \cos(2\beta) m_Z^2$ and $D_R = s^2_W \cos(2\beta) m_Z^2$. Throughout this article, we shall use universal slepton mass parameters $M_\slepton$ for all generations and assume $M_L^2 = M_R^2 (\equiv M_{\slepton}^2)$. The diagonal entries of \eq{eq:Mslep} will be denoted as $m_{\slepton_L}^2=M_L^2 +m_l^2 + D_L$ and $m_{\slepton_R}^2=M_R^2 + m_l^2 + D_R$. Note that due to the small mixing for the first and second generations of sleptons, these approximately coincide with the mass eigenvalues. For the third generation, i.e., the stau, the mixing is larger. Thus, the lightest stau will be the lightest slepton.

Due to the smallness of the Yukawa couplings, the most relevant interactions for collider phenomenology are those associated with the gauge couplings. The couplings to gauge bosons provide the dominant terms for chargino and neutralino production (for a more extended discussions see Ref.~\cite{Liu:2020ctf}). Considering the decays of these neutralinos and charginos, their interactions with the Higgs, gauge bosons and sleptons are the most important ones~(again proportional to the electroweak gauge couplings), and will be described in more detail below. 

Beyond the electroweak interacting particles, the strongly interacting sector plays a relevant role in the phenomenology of low energy supersymmetry. First we stress that the value of the squark masses plays a role in the electroweakino production cross section, due to a $t$-channel contribution that interferes destructively with the gauge boson induced one (see, for instance, Ref.~\cite{Liu:2020ctf}). This implies that the bounds on the electroweakinos and sleptons can be weaker than the ones typically presented by the LHC collaborations since those results usually assume very large squark masses, suppressing destructive interference effects in the electroweakino production-cross-section calculation. In this article, we shall assume that the squark masses and the gluino mass are all of the order of 2.5\,TeV, and thus beyond the current direct reach of the LHC. For these values of the squark mass parameters and in the range $\tan\beta \sim 50 - 70$ we will focus on, a value of the up-type quark $A$-term of $A_t \sim 3.5\,$TeV will then lead to the proper mass of the SM-like Higgs boson, $m_h \sim 125\,$GeV. Finally, we assume large values of the CP-odd Higgs mass, of the order of the squark masses, $m_A \sim 2.5\,$TeV.

\section{Phenomenology in the Compressed Region \label{sec:analytical}} 

We shall concentrate on the regions of parameter space allowed by current electroweakino searches, which implies that the neutralino mass parameters $M_1$, $M_2$ and $\mu$ are of the order of or above a few hundred GeV, and thus significantly larger than $m_Z$. Therefore, the mixings in the neutralino sector are suppressed. Requiring a viable DM candidate further constrains our parameter space. Winos and Higgsinos in the few-hundred GeV mass range annihilate too efficiently to explain the observed DM relic density. Hence, we focus on the region of the parameter space where $|M_1| \lesssim |M_2| \lesssim |\mu|$, i.e., where the lightest neutralino is ``Bino-like'' $\chione\approx\widetilde{B}$, the next-to-lightest neutralino is ``Wino-like" $\chitwo\approx\widetilde{W}^0$, and the remaining neutralinos are ``Higgsino-like" $\chithree,\chifour\approx (\widetilde{H}_u^0\pm\widetilde{H}_d^0)/\sqrt{2}$. For a Bino-like DM candidate $\chione$, it is well known that $\chione$ pair-annihilation is not efficient enough to deplete the DM relic density to the right amount~\cite{Ellis:1999mm,Buckley:2013sca,Cabrera:2016wwr}. As we will discuss, Bino-Wino co-annihilation is the most straightforward possibility to obtain the observed DM relic density for a Bino-like $\chione$ while explaining the observed value of the muon's magnetic dipole moment and satisfying collider constraints. For the mass range of the DM candidate we will be most interested in, $m_\chione \sim 100-400\,$GeV, we find that the observed relic density is typically obtained for values of the mass splitting between the Wino-like $\chitwo$ and $\chione$ of $(\massplit) \sim 10-30\,$GeV. Throughout, we will refer to the region of parameter space featuring such mass splitting as the {\it compressed region}. As we will see, not only are we able to obtain a consistent result for the measured value of the muon's magentic dipole moment and satisfy all current direct detection constraints in the compressed region, but the Wino-like $\chitwo$~(which can be copiously produced at the LHC) generically has a significant radiative decay $(\chitwo\rightarrow\chione+\gamma)$ branching ratio, giving rise to a potentially interesting collider signal at the LHC. 

\subsection{Dark Matter Constraints \label{sec:relicDensity}} 

The observed DM relic density is $\Omega_{\rm DM} h^2 \simeq 0.12$~\cite{Planck:2018vyg}, where $h \equiv H_0/(100\,{\rm km/s/Mpc})$ parameterizes the Hubble constant and $\Omega_{\rm DM} \equiv \rho_{\rm DM}/\rho_c$ is the DM density in units of the critical density. In order for $\chione$ to be a viable DM candidate, its relic density, which we denote as $\Omega_{\chione} \equiv \rho_{\chione}/\rho_c$, should match the observed value $\Omega_{\rm DM}$. For DM produced via standard thermal freeze-out~(FO), the relic density and the effective (thermally averaged) annihilation cross section at freeze out, $\left\langle \sigma v \right\rangle_{\rm FO}$, are approximately related as
\begin{equation} \label{eq:relic_density}
    \Omega_{\chione} h^2 \sim 0.1 \times \frac{3 \times 10^{-26}\,{\rm cm}^3/{\rm s}}{\left\langle \sigma v \right\rangle_{\rm FO}} \;.
\end{equation}
Since at the time of freeze-out (parameterized by the temperature $x \equiv m_{\chione}/T$) the DM candidate is usually non-relativistic (typically, $x^{\rm FO} \sim 20$), one often expands $\left\langle \sigma v \right\rangle_{\rm FO}$ as
\begin{equation} \label{eq:annxsec_NR}
    \left\langle \sigma v \right\rangle_x = a + 6 b / x + \mathcal{O}(x^{-2})\;.
\end{equation}

In our scenario, the DM candidate $\chione$ is Bino-like, $\chione \sim \widetilde{B}$. For a pure Bino with mass in the few-hundred GeV range, $\left\langle \sigma v \right\rangle_{\rm FO}$ tends to be so small that the resulting relic density would overclose the Universe, $\Omega_{\chione} h^2 \gg 0.1$. The dominant annihilation mechanism for a pure Bino is into pairs of SM fermions, $\chione\chione \to f \bar{f}$, mediated via $t$-channel exchange of the associated sfermion $\widetilde{f}$. The corresponding amplitudes are inversely proportional to the square of the mass of the sfermion in the $t$-channel. For any $\chione\chione \to f \bar{f}$ process, three amplitudes contribute to the cross section. Two of the diagrams are mediated by $t$-channel exchange of the left-handed or the right-handed sfermion, respectively. Both of these amplitudes are $p$-wave and hence contribute to $b$ in the non-relativistic expansion as written in \eq{eq:annxsec_NR}. Hence, the contribution of these diagrams to the total cross section at freeze-out is suppressed by $1/x^{\rm FO} \sim 1/20$. The third diagram involves left-right mixing of the slepton in the $t$-channel; the corresponding amplitude is $s$-wave and contributes to $a$ in \eq{eq:annxsec_NR}, however, it is suppressed by the small left-right mixing of sfermions.

 We consider a region of the MSSM parameter space where the squarks are much heavier than the sleptons, hence, $\chione\chione \to l \bar{l}$ processes will be more important than $\chione\chione \to q \bar{q}$. The largest contribution to the annihilation cross section of a pure Bino would come thus from the third-generation sleptons (i.e., the staus, $\widetilde{\tau}$) since they exhibit the largest left-right mixing among the sleptons [cf. \eq{eq:Mslep}]. This mechanism can give rise to sufficiently large effective annihilation cross section to produce the observed DM density if the sleptons are light enough, $m_{\tilde{l}} \lesssim 150\,$GeV~\cite{Pierce:2013rda,Fukushima:2014yia}. However, for such light sleptons (requiring an even lighter Bino-like DM candidate), it is challenging to avoid direct detection and collider constraints and simultaneously explain the observed value of the muon's magnetic dipole moment, hence, we will not explore this option further in this work.

The effective annihilation cross section of a Bino-like DM candidate can be enhanced to the required value $\left\langle \sigma v \right\rangle_{\rm FO} \sim 3 \times 10^{-26}\,{\rm cm}^3/{\rm s}$ by a number of different mechanisms. First, a sufficiently large Higgsino-admixture can make annihilation of $\chione$'s into pairs of Higgs bosons or SM fermions mediated by any of the MSSM's Higgs bosons or the $Z$-boson in the $s$-channel sufficiently effective to lead to the observed relic density~\cite{Arkani-Hamed:2006wnf}. However, it is difficult to satisfy current direct detection constraints in this ``well-tempered'' neutralino scenario.\footnote{Note that a ``well-tempered'' Bino-Wino admixture is very challenging to realize since the Bino and the neutral Wino do not directly mix, but only via the Higgsinos, see \eq{eq:mn_matrix}.} Second, one can use the same diagrams as the well-tempered neutralino scenario and compensate a smaller Higgsino fraction of $\chione$ by resorting to resonant $s$-channel annihilation in the ``$Z/h/A$-funnels'' solutions~\cite{Han:2013gba,Cabrera:2016wwr}. The smaller Higgsino fraction alleviates direct detection constraints, but such a solution requires the mass of the DM candidate to be tuned to half of the mass of the $s$-channel mediator. The $Z/h/A$-funnels solutions are challenging to realize while simultaneously explaining the observed value of the muon's magnetic dipole moment and avoiding current LHC constraints~\cite{Baum:2021qzx}. 

Third, the effective annihilation cross section of the DM candidate can be enhanced by co-annihilation with sleptons, heavier neutralinos and/or charginos~\cite{Ellis:1998kh,Ellis:1999mm,Buckley:2013sca,Han:2013gba,Cabrera:2016wwr,Baker:2018uox, Yanagida:2019evh}. The basic mechanism of co-annihilation works as follows~\cite{Griest:1990kh}: if a second (or multiple additional) states $\tilde{\chi}_{i}$ exist with masses $m_{\tilde{\chi_i}}$, and if there are processes which can effectively convert $\chione \leftrightarrow \tilde{\chi}_i$, the number density of the $\tilde{\chi_i}$ and of $\chione$ at temperature $T$ will be related by a Boltzmann-factor, $n_{\tilde{\chi}_i} \sim n_{\chione} \exp\left[ - \left( m_{\tilde{\chi}_i}-m_{\chione} \right) / T \right] $. If the annihilation cross section of processes involving the $\tilde{\chi}_i$ are large compared to $\chione$'s annihilation processes, and if the mass gap $( m_{\tilde{\chi}_i}-m_{\chione} )$ is not too large, the effective annihilation cross section for \eq{eq:relic_density} can be dominated by the annihilation cross section of the $\tilde{\chi}_i$'s multiplied with a Boltzmann-suppression factor accounting for the difference in the number densities $n_{\tilde{\chi}_i}$ and $n_{\chione}$. 

In this work, we will focus on the region of parameter space where the observed relic density $\Omega_{\rm DM} h^2 \simeq 0.12$ of a Bino-like DM candidate is achieved via co-annihilation with the Wino-like neutralino and chargino states: if the Wino-like neutralino and chargino states have masses not much larger than those of the Bino-like DM candidate, the right relic density of the Bino-like DM candidate can be realized by co-annihilation with the Wino-like states which have much larger annihilation cross sections. The mass of the Bino-like neutralino is controlled by the associated supersymmetry-breaking mass parameter, $m_{\chione} \approx |M_1|$. The Wino-like neutralino and charginos are approximately mass degenerate, $m_{\chitwo} \approx m_{\chaone} \approx |M_2|$. In the region of parameter space of $m_{\chione} \sim 100-500\,$GeV we will focus on here, the observed relic density is obtained for a mass splitting $(\massplit) \sim 10-30\,$GeV. 

 The null results from DM direct detection experiments impose strong constraints on the parameter space. The spin-independent (SI) scattering between $\chione$ and the nuclei is mediated by $t$-channel exchange of CP-even Higgs bosons, the amplitude of which is controlled by the Higgsino components of $\chione$. The SI scattering amplitude between $\chione$ and the nuclei is proportional to~\cite{Huang:2014xua,Han:2018gej,Baum:2021qzx}:
\begin{equation} \label{eq:SIDD}
    \mathcal{M}_p^{\rm SI}\propto \frac{v}{\mu^2}\left[2\frac{(M_1+\mu\sin 2\beta)}{m_h^2} - \frac{\mu \cos2\beta \tan\beta}{m_H^2} \right] \approx \frac{v}{\mu^2}\left[2\frac{(M_1+ 2 \mu / \tan \beta)}{m_h^2} + \frac{\mu \tan \beta}{m_H^2} \right] ;
\end{equation}
the approximation holds for moderate to large values of $\tan\beta$ where $\sin 2\beta \to 2/\tan\beta$ and $\cos2\beta \to -1$. If $M_1$ and $\mu$ have opposite signs, the two terms contributing to the amplitude mediated by the lighter Higgs boson [the first term in the square bracket in \eq{eq:SIDD}] partially cancel. Furthermore, if $(M_1+2\mu/\tan\beta)$ and $(\mu\tan\beta)$ have opposite signs, which due to the $\tan\beta$ suppression/enhancement of the respective terms generally corresponds to $M_1$ and $\mu$ having opposite signs, the contributions from the light ($h$) and heavy ($H$) Higgs exchange channels interfere destructively. As discussed in the introduction, for $(M_1\times \mu) <0$, this suppression of the SI cross section compared to the $(M_1\times \mu)>0$ case allows the MSSM to satisfy direct detection constraints without the need of large values of $|\mu|$. As we will see in \sect{sec:Computations}, simply choosing $M_1$ and $\mu$ to have opposite signs suffices to suppress the SI cross sections to values below current experimental constraints for a sizable region of parameter space, see also Ref.~\cite{Baum:2021qzx} for a discussion. The limit of $M_1$, $\mu$, $\tan\beta$, and $m_H$ being arranged such that $\mathcal{M}_p^{\rm SI} \to 0$ is known as the ``generalized blind spot''~\cite{Huang:2014xua}; if future direct detection searches do not find DM, then only-fine-tuned blind-spot solutions will remain allowed in the MSSM for a Bino-like DM candidate with few-hundred GeV mass, unless $|\mu|$ takes values well above a TeV implying significant finetuning of the electroweak scale.

Let us also briefly comment on spin-dependent (SD) scattering. The corresponding cross section is dominated by $Z$ exchange; at moderate to large $\tan\beta$, the SD scattering amplitude is proportional to (see for example, Ref.~\cite{Baum:2021qzx}):
 \begin{equation}
     \mathcal{M}^{\rm SD}\propto \left( \frac{v}{\mu} \right)^2\cos 2\beta \;.
\end{equation}
Note that there is no cancellation of different contributions to the cross section, the SD cross section is simply controlled by the absolute size of the Higgsino admixtures to the DM candidate. Current bounds from direct detection experiments place much weaker constraints on the DM scenarios in the MSSM than SI bounds. However, if future DM searches do not find DM, then SD searches will become a powerful probe of blind-spot type solutions that remain allowed by SI searches, see, for example, Refs.~\cite{Cohen:2011ec,Baum:2017enm} for discussions.

\subsection{Muon Magnetic Moment \label{sec:gm2}}

A major point of interest for our analysis is the recent $\gmtwo$ result reported by the muon g-2 collaboration at Fermilab~\cite{Muong-2:2021ojo}. With the Fermilab result, the tension between SM predictions and experimental measurements, $\Delta a_\mu = a_\mu^{\text{exp}} - a_\mu^{\text{SM}} = \left( 25.1 \pm 5.1 \right) \times 10^{-10}$, has grown to a significance of $\sim 4.2\,\sigma$. The MSSM provides an interesting explanation for this discrepancy, as MSSM particles provide contributions to the muon-photon vertex. We are interested in studying the case where the MSSM corrections resolve the tension between theory and experiment, and so we consider these corrections to be the only BSM contributions to the theoretical prediction for \gmtwo. The two leading MSSM contributions are the chargino-sneutrino loop ($a_\mu^{\tilde{\chi}^{\pm}}$) and the Bino-slepton loop ($a_\mu^{\tilde{\chi}^0}$). They are approximately given by~\cite{Barbieri:1982aj,Ellis:1982by,Kosower:1983yw,Moroi:1995yh,Carena:1996qa,Feng:2001tr,Martin:2001st,Badziak:2019gaf}
\begin{eqnarray}
  a_\mu^{\tilde{\chi}^{\pm}}  &\approx& \frac{ \alpha m_\mu^2 \mu M_2 \tan \beta}{4\pi \sin^2 \theta_W m_{\tilde \nu_\mu}^2} \left[ \frac{f_{\tilde{\chi}^\pm} (M_2^2/m_{\tilde \nu_\mu}^2 ) - f_{\tilde{\chi}^\pm}(\mu^2/m_{\tilde \nu_\mu}^2)}{M_2^2 - \mu^2} \right], \label{eq:achargino}
    \\
   a_\mu^{\tilde{\chi}^0} &\approx&  \frac{ \alpha m_\mu^2 M_1 \left( \mu \tan \beta - A_\mu \right)}{4\pi \cos^2 \theta_W \left( m_{\tilde l_R}^2 - m_{\tilde l_L}^2 \right)} \left[ \frac{f_{\tilde{\chi}^0}(M_1^2/m_{\tilde l_R})}{m_{\tilde l_R}^2} - \frac{f_{\tilde{\chi}^0}(M_1^2/m_{\tilde l_L})}{m_{\tilde l_L}^2} \right], \label{eq:aneutralino}
\end{eqnarray}
where the two functions are
\begin{eqnarray}
    f_{\tilde{\chi}^\pm}(x) &=& \frac{x^2 - 4x + 3 + 2 \ln x}{\left(1-x\right)^3} \;,
    \\
    f_{\tilde{\chi}^0}(x) &=& \frac{x^2 -1 - 2x \ln x}{\left(1-x\right)^3} \;,
\end{eqnarray}
both are defined at $x=1$ to be continuous, $f_{\tilde{\chi}^\pm}(1) = -2/3$ and $f_{\tilde{\chi}^0}(1) = -1/3$. Because we have assigned the left-handed and right-handed sleptons equal soft masses ($M_{L} = M_{R}$), the physical masses will be quite close. It appears that this will cause the contribution from the Bino-slepton loop, $a_\mu^{\tilde{\chi}^0}$, to become infinite, but the cancellation of the two factors in the square bracket of \eq{eq:aneutralino} causes $a_\mu^{\tilde{\chi}^0}$ to remain continuous in the limit $m_{\tilde l_R} \to m_{\tilde l_L}$. \eq{eq:achargino} and \eq{eq:aneutralino} capture the leading MSSM contribution to the muon's anomalous magnetic dipole moment, $a_{\mu}^{\text{MSSM}} \approx a_\mu^{\tilde{\chi}^\pm} + a_\mu^{\tilde{\chi}^0}$. Note that for our numerical results, we employ \verb|MicrOMEGAS| to calculate $a_{\mu}^{\text{MSSM}}$, including all MSSM contributions at one-loop order. The leading log two-loop contributions were evaluated in, e.g., Ref~\cite{Martin:2001st}. The modification factor with respect to the one-loop results is roughly 8\% for the superpartner mass scale considered in this work, well below the uncertainty of the current observed value of $\Delta a_\mu$ ($\sim 20\%$). Hence, one-loop order contributions are sufficient for the purpose of this work.

For $|M_1| \sim |M_2|$, the Wino-sneutrino contribution $a_\mu^{\tilde{\chi}^\pm}$ will be the dominant contribution to $a_\mu^{\rm MSSM}$ since $\sin^2\theta_W \sim 0.2$ while $\cos^2\theta_W \sim 0.8$. Thus, in order to generate a correction which resolves the $\Delta a_\mu$ problem, we require $a_\mu^{\tilde{\chi}^\pm}$ to be positive, and therefore $(M_2 \times \mu) > 0$. On the other hand, we see from \eq{eq:aneutralino} that the diagrams will add constructively for $(M_1 \times \mu) > 0$ or destructively for $(M_1 \times \mu) < 0$, leading to an $\mathcal O(40)\,\%$ change of $a_\mu^{\rm MSSM}$ when one flips the sign of $M_1$. In general, if one wants a specific value of $a_\mu^{\rm MSSM}$, one must compensate the interference with larger $\mu$ or $M_2$ for $(M_1 \times \mu) > 0$, and vice versa.

As we will see in \sect{sec:Computations}, we find MSSM contributions explaining the measured value of $a_\mu$ in the approximate range $|\mu| \sim 400 - 1000\,$GeV; the precise values depends on the choice of $M_1, M_2, \tan\beta, A_\mu,$ and $M_{\tilde l}$, but always require $(M_2 \times \mu) > 0$. If we require a DM candidate with the observed DM relic density compatible with current direct detection experiments, the sign-combination $(M_1 \times \mu) < 0$ opens up much larger regions of parameter space than $(M_1 \times \mu) > 0$ due to the cancellations in the spin-independent direct detection cross section discussed in \sect{sec:relicDensity}. 

\subsection{Decays from $\chitwo$ to $\chione$}
\label{sec:decay}

As summarized above, simultaneously explaining the measured value of the muon's magnetic dipole moment and providing a DM candidate compatible with current direct detection bounds without exacerbating electroweak finetuning problems motivates considering a region of parameter space where the lightest neutralino ($\chione$) is Bino-like, where the second-lightest neutralino ($\chitwo$) and the lightest chargino ($\chaone$) are Wino-like with masses not much larger than $\chione$, and where $(M_1 \times \mu) < 0$ and $(M_2 \times \mu) > 0$. The observed relic density for $\chione$ is achieved via Bino-Wino co-annihilation in the ``compressed region'', $(\massplit) \sim 10-30\,$GeV. As we will see in this section, these relative signs, which imply $(M_1 \times M_2) < 0$, have important implications on the decay patterns of the Wino-like neutralino in the compressed region. The analytical approximations for the decay modes of the second-lightest neutralino, which we discuss here, will serve as guidance for our subsequent numerical analysis.

\begin{figure}
\centering
\begin{subfigure}{0.3\textwidth}
\includegraphics[width=\textwidth]{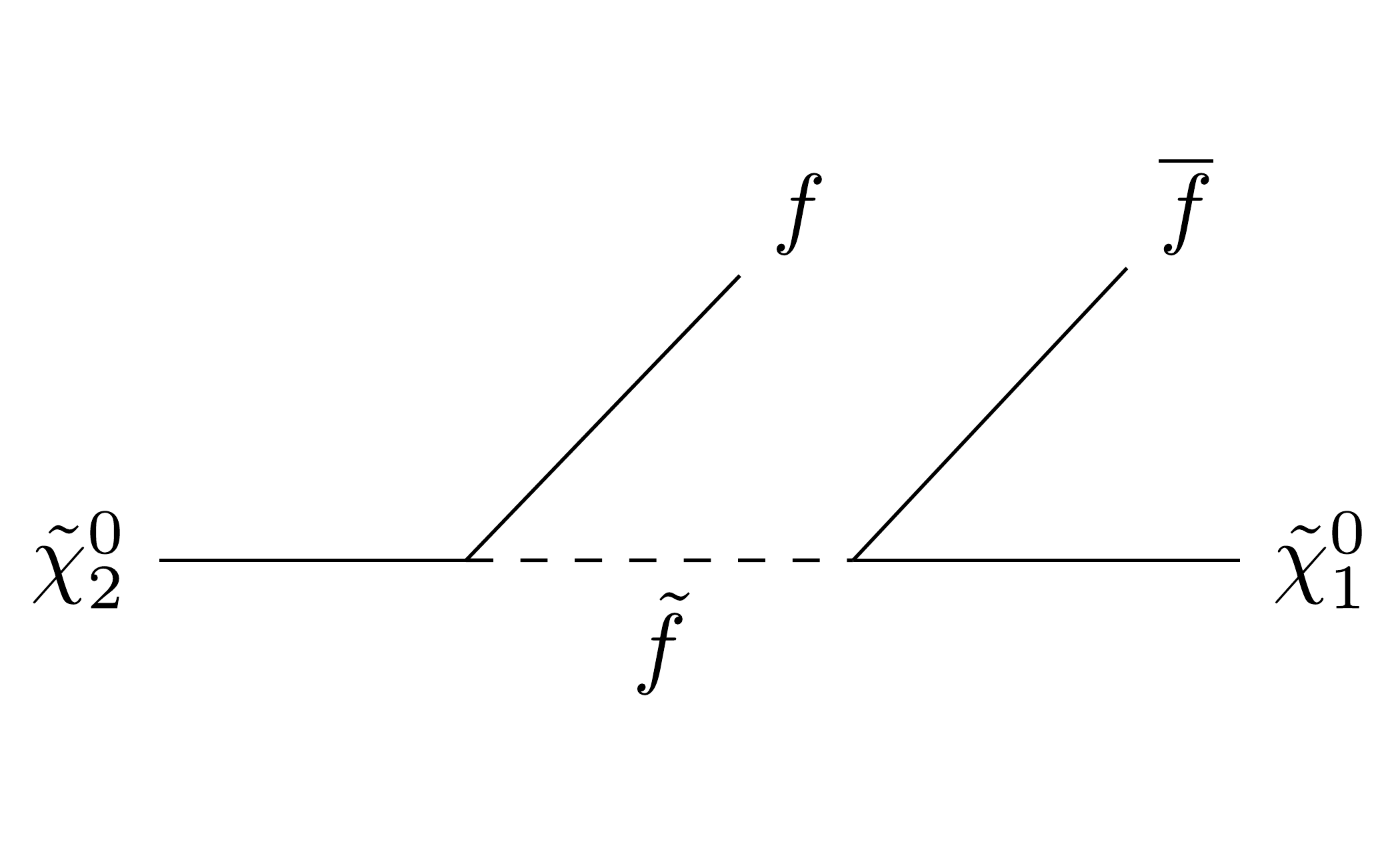}
\caption{}
\label{fig:tree_sf}
\end{subfigure}
~\begin{subfigure}{0.3\textwidth}
\includegraphics[width=\textwidth]{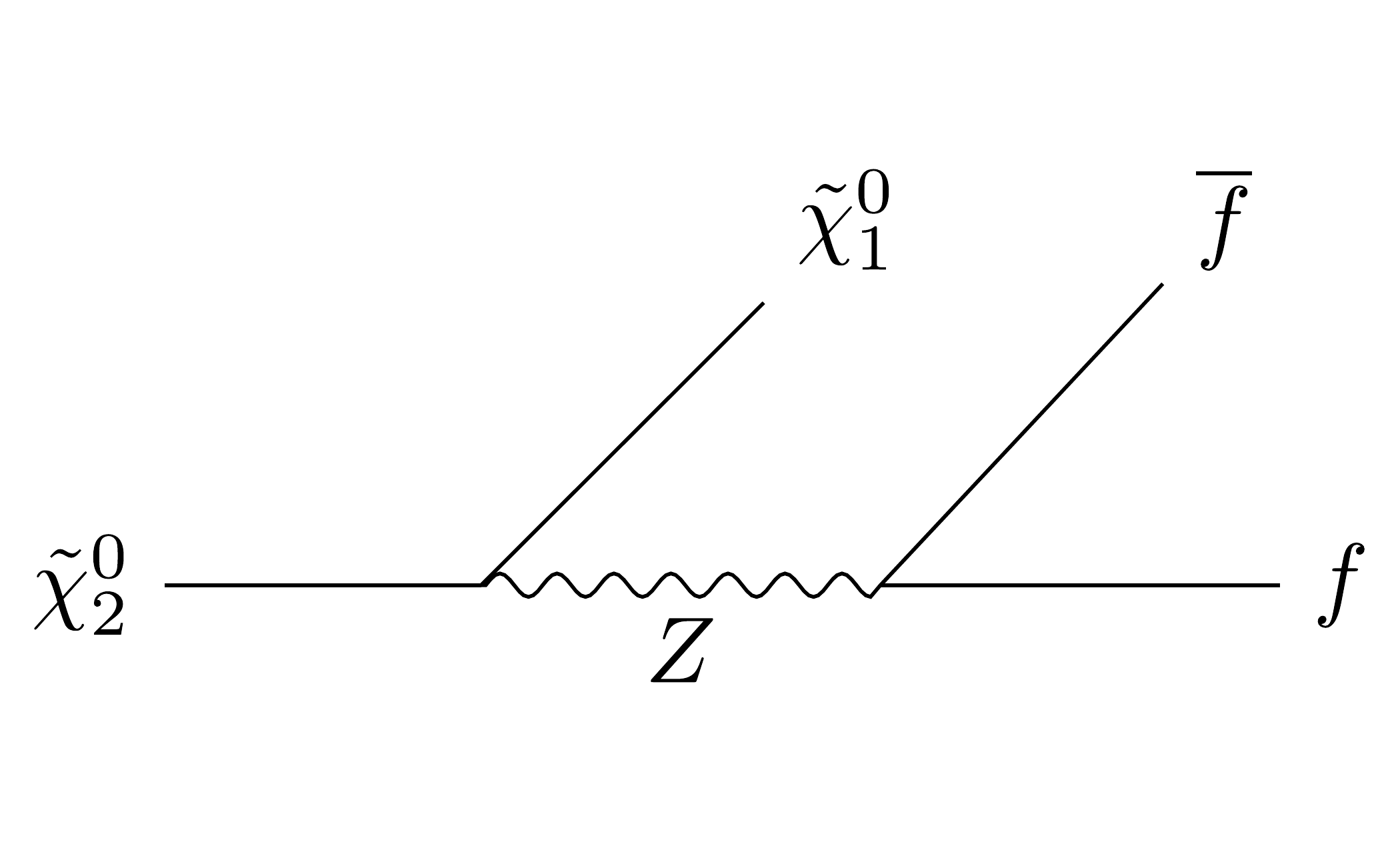}
\caption{}
\label{fig:tree_Z}
\end{subfigure}
~\begin{subfigure}{0.32\textwidth}
\includegraphics[width=\textwidth]{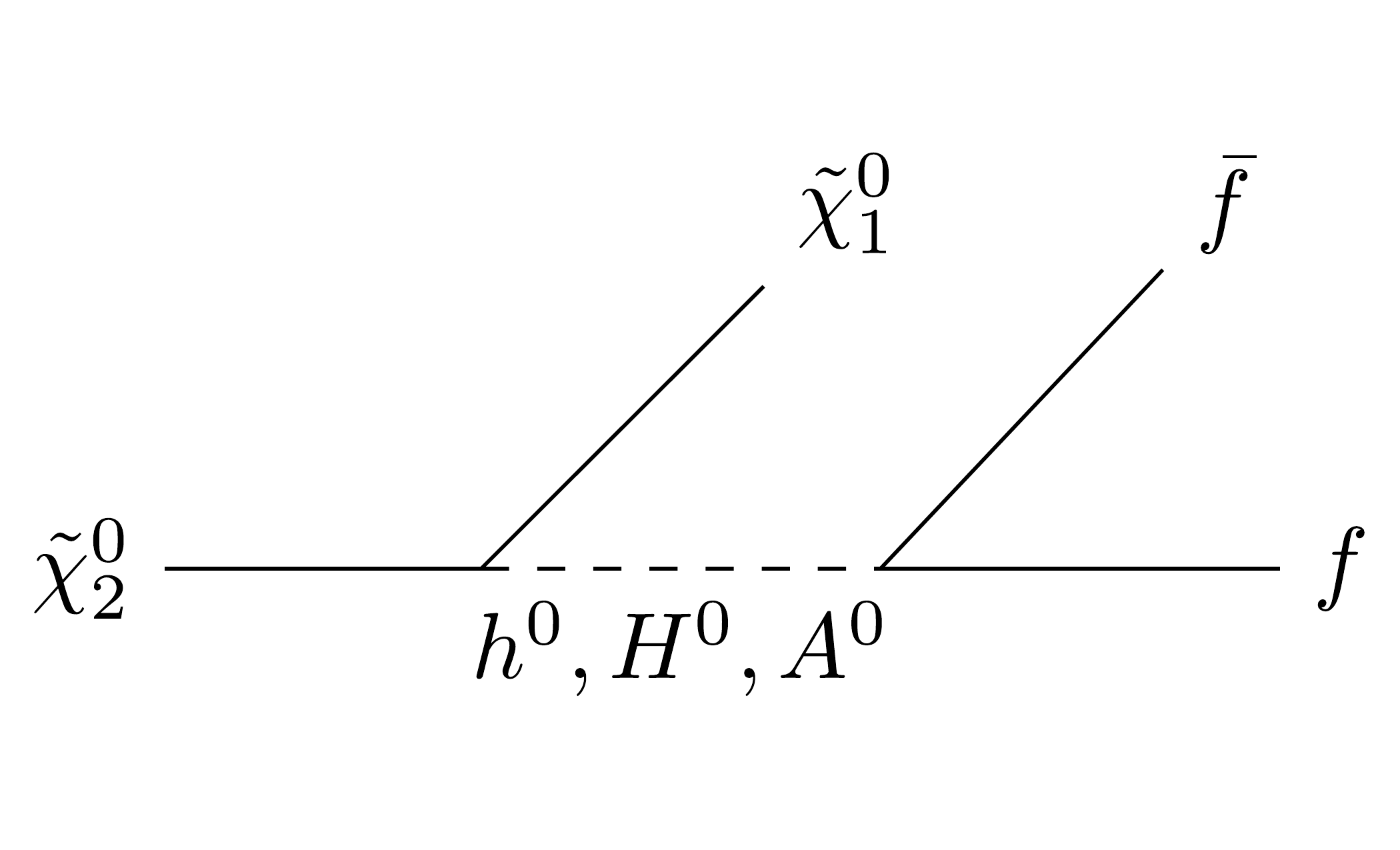}
\caption{}
\label{fig:tree_h}
\end{subfigure}
\caption{Tree-level decays of the second-lightest neutralino $\chitwo$ to the lightest neutralino $\chione$ and a pair of SM fermions ($f + \bar{f}$).}
\label{fig:chitwo-to-chione-ff}
\end{figure}

All the decay modes for $(\chitwo\rightarrow\chione+X)$ are kinematically suppressed in the compressed region. To parameterize the kinematic suppression, we define the ``mass splitting parameter''
\begin{equation}
    \varepsilon\equiv \frac{m_{\chitwo}}{m_{\chione}}-1.
\end{equation}
Tree-level decays (illustrated in \fig{fig:chitwo-to-chione-ff}) are suppressed as $\Gamma(\chitwo\to\chione+ f\bar{f}) \propto \varepsilon^5$, while the radiative decays (illustrated in \fig{fig:sf_f_loop}--\fig{fig:W_chi_loop}) are suppressed as $\Gamma(\chitwo\to\chione+\gamma) \propto \varepsilon^3$ \cite{Ambrosanio:1996gz}. Therefore, radiative decays play an important role in the compressed region. As we shall see, the radiative decay width is enhanced if $M_1$ has a negative sign relative to $M_2$; a similar effect was also observed in Ref.~\cite{Baer:2002kv}. Recall that we encode the signs of the neutralino masses obtained from the diagonalization of the mass matrix in $\xi_i$, hence ($M_1 \times M_2) < 0$ corresponds to ($\xi_1 \times \xi_2) = -1$. 

The radiative decay width is~\cite{Haber:1988px}
\begin{equation}
   \Gamma(\chitwo\rightarrow\chione+\gamma)=\frac{g^2_{\chitwo\chione\gamma} \left(m_\chitwo^2-m_\chione^2 \right)^3}{8\pi m_\chitwo^5} \;,
\end{equation}
where $g_{\chitwo\chione\gamma}$ is the total effective coupling contributed by three types of triangle loops as discussed below:

\begin{figure}
        \centering
        \begin{subfigure}{0.3\textwidth}
        \includegraphics[width=\textwidth]{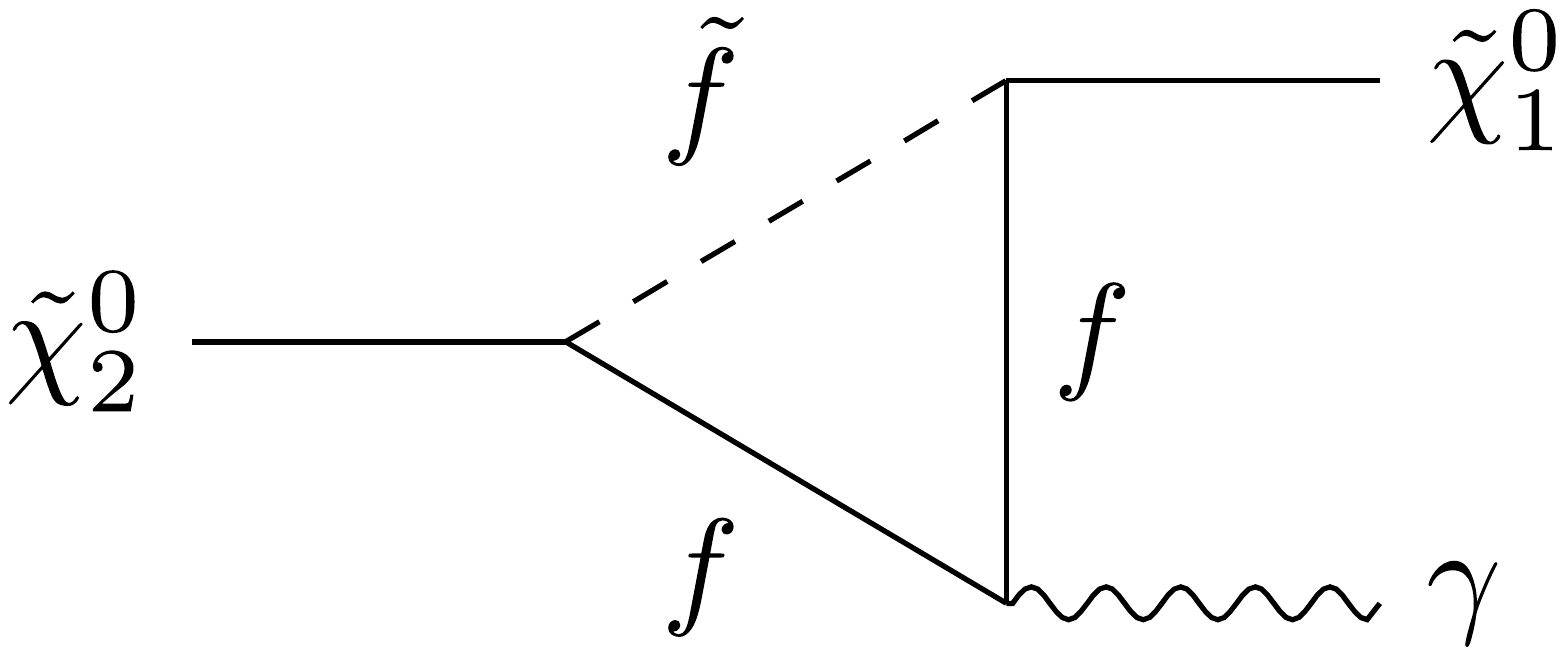}
        \end{subfigure}
        ~~~\begin{subfigure}{0.3\textwidth}
        \includegraphics[width=\textwidth]{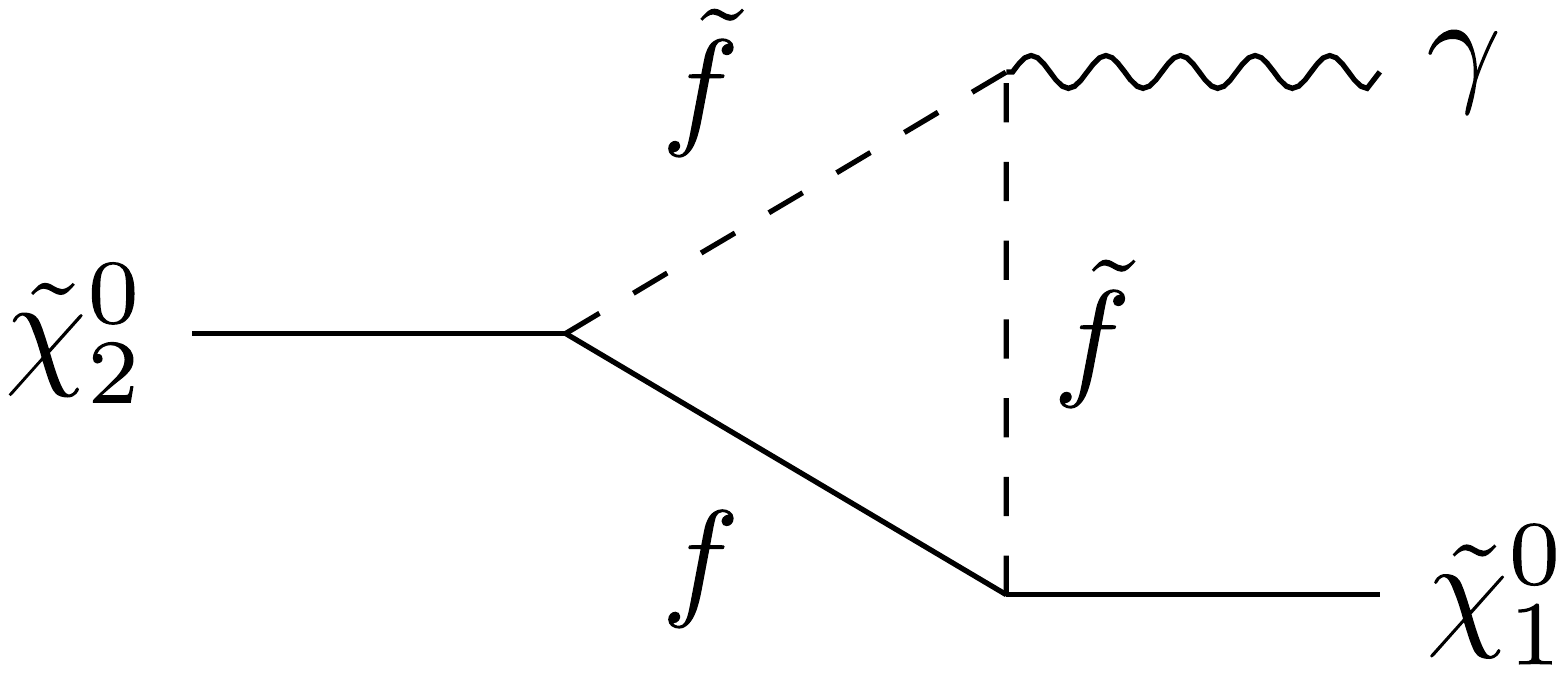}
        \end{subfigure}
        \caption{Sfermion-fermion ($\sfermion-f$) triangles contributing to the radiative decay.}
        \label{fig:sf_f_loop}
\end{figure}

\begin{enumerate}
     \item \bm{$\sfermion-f$} \textbf{triangle:} The corresponding diagrams are shown in \fig{fig:sf_f_loop}. Calculating the total matrix element of these diagrams containing a $\sfermion-f$ triangle, and extracting the effective coupling, we get 
     \begin{align} 
        g_{\sfermion/f} = \frac{e g^2  m_{\chitwo}}{32 \pi^2} \sum_f Q_f C_f & \left\{ \left( G_L F_R - G_R F_L \right) \left[ (\xi_1\times\xi_2) m_{\chitwo} \left( I_2 - K \right) -  m_{\chione}K \right] \right. \nonumber \\
        & \left. \phantom{m_{\chitwo}} + \xi_1 m_f \left( G_L F_L - G_R F_R \right) I \right\} \;, \label{eq:sf_f_loop}
    \end{align}
    where $Q_f$, $C_f$ and $m_f$ are the electric charge, color charge and the mass of the fermion in the loop, respectively. $F_{L(R)}$ and $G_{L(R)}$ are the couplings of the incoming and outgoing neutralinos to the particles in the loop. For up-type fermions/sfermions, the relevant combinations in \eq{eq:sf_f_loop} are given by
    \begin{equation}
       \hspace*{-4cm}G_L F_R -G_R F_L=\mathbf{N}_{1}^{-} \mathbf{N}_{2}^{-}+ 4 T_{3u} Q_u \tan\theta_W \left( \mathbf{N}_{11}\mathbf{N}_{2}^{-}+\mathbf{N}_{21}\mathbf{N}_1^{-} \right),
       \label{eq:glfrgrfl_f}
   \end{equation}
   \begin{align}
       G_L F_L - G_R F_R &= -\frac{2m_u}{m_W \sin\beta} \bigg[ \mathbf{N}_{14}\left(T_{3u} \mathbf{N}_{2}^{-}+Q_u \tan\theta_W \mathbf{N}_{21}\right) \label{eq:glflgrfr_f} \\
       & \quad -\mathbf{N}_{24}\left(T_{3u}\mathbf{N}_{1}^{-}+Q_u\tan\theta_W \mathbf{N}_{11}\right)-Q_u\tan\theta_W\left(\mathbf{N}_{24}\mathbf{N}_{11}-\mathbf{N}_{14}\mathbf{N}_{21}\right)\bigg] \;, \nonumber
    \end{align}
    where $T_{3u}$ is the isospin of the fermion in the loop, and $Q_u$ is the electric charge of the fermion in units of $e$. The $\mathbf{N}_{ij}$ are the matrix elements of the neutralino mixing matrix defined in \sect{sec:spec}, and $\mathbf{N}_{i}^{\pm}\equiv \mathbf{N}_{i2}\pm \mathbf{N}_{i1}\tan\theta_W$. The results for down-type fermions/sfermions corresponding to \eq{eq:glfrgrfl_f} and \eq{eq:glflgrfr_f} can be obtained by making the following replacements: $(m_u,Q_u,T_{3u})\rightarrow (m_d, Q_d, T_{3d})$, $\sin\beta\rightarrow\cos\beta$, and $(\mathbf{N}_{14},\mathbf{N}_{24})\rightarrow (\mathbf{N}_{13}, \mathbf{N}_{23})$. 
    
    \eq{eq:glflgrfr_f} indicates that $(G_L F_L-G_R F_R)$ is suppressed by the SM fermion masses and neutralino mixing, while $(G_L F_R-G_R F_L)\approx \mathbf{N}_{11}\mathbf{N}_{22}\tan\theta_W(-1+4T_3 Q)$ is neither suppressed by the fermion masses nor by the neutralino mixing. Hence we will focus on the first line of \eq{eq:sf_f_loop}. Since $\chione$ and $\chitwo$ are Majorana fermions, the fermion flow in \fig{fig:sf_f_loop} can be either  preserved or violated. This gives rise to the dependence on the signs of the neutralino masses $\xi_1$ and $\xi_2$ in \eq{eq:sf_f_loop}. The full expressions of the loop integrals $I_2$ and $K$ are given in \app{app:radiative_decay}. Expanding $(I_2-K)$ and $K$ in terms of the mass splitting parameter $\varepsilon$, we get
    \begin{eqnarray}
        I_2-K &=& \int_0^1 dx~\frac{1}{\mathcal{D}} x \left(x-1\right) +\mathcal{O}(\varepsilon) \;, \\
        K &=& \int_0^1 dx~\frac{1}{\mathcal{D}} x \left(x-1\right) +\mathcal{O}(\varepsilon) \;.
    \end{eqnarray}
   The parameter $\mathcal{D}$ is defined as
    \begin{eqnarray}
        \mathcal{D} &\equiv& 2\left[ x m_{f}^2 +(1-x) m_{\sfermion}^2-x (1-x)m_{\chione}^2\right] \;,
        \label{eq:def_D}
    \end{eqnarray}
    where $m_\sfermion$ and $m_f$ are the masses of the sfermion and fermion in the loop, respectively. At leading order in $\varepsilon$, $I_2 - K = K + \mathcal O(\varepsilon)$. Hence, the effective coupling $g_{\sfermion/f}$, \eq{eq:sf_f_loop}, is enhanced if ($\xi_1 \times \xi_2) = -1$, corresponding to $(M_1 \times M_2) < 0$.
     
    \item \bm{$H^\pm/G^\pm-\chak$} \textbf{triangle:} The diagrams are shown in \fig{fig:H_chi_loop}. The effective coupling has the same form as that of the $\sfermion-f$ triangle, with $Q_f=1$ and $C_f=1$. However, since both the charged Higgs and the Higgsino-like chargino are heavy in the parameter space under analysis, the contribution from these types of triangles tends to be highly suppressed. 
    
    \begin{figure}
        \centering
        \begin{subfigure}{0.3\textwidth}
        \includegraphics[width=\textwidth]{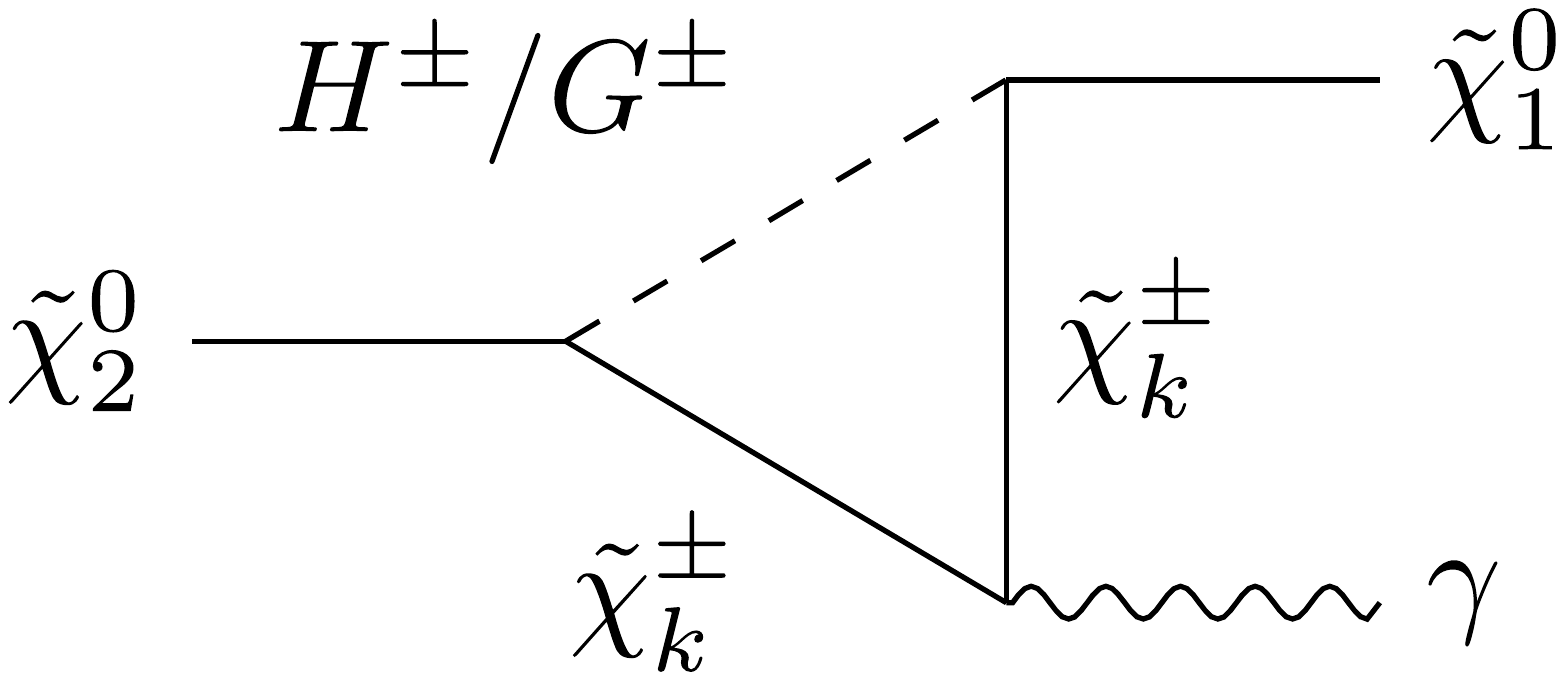}
        \end{subfigure}
        ~~~\begin{subfigure}{0.3\textwidth}
        \includegraphics[width=\textwidth]{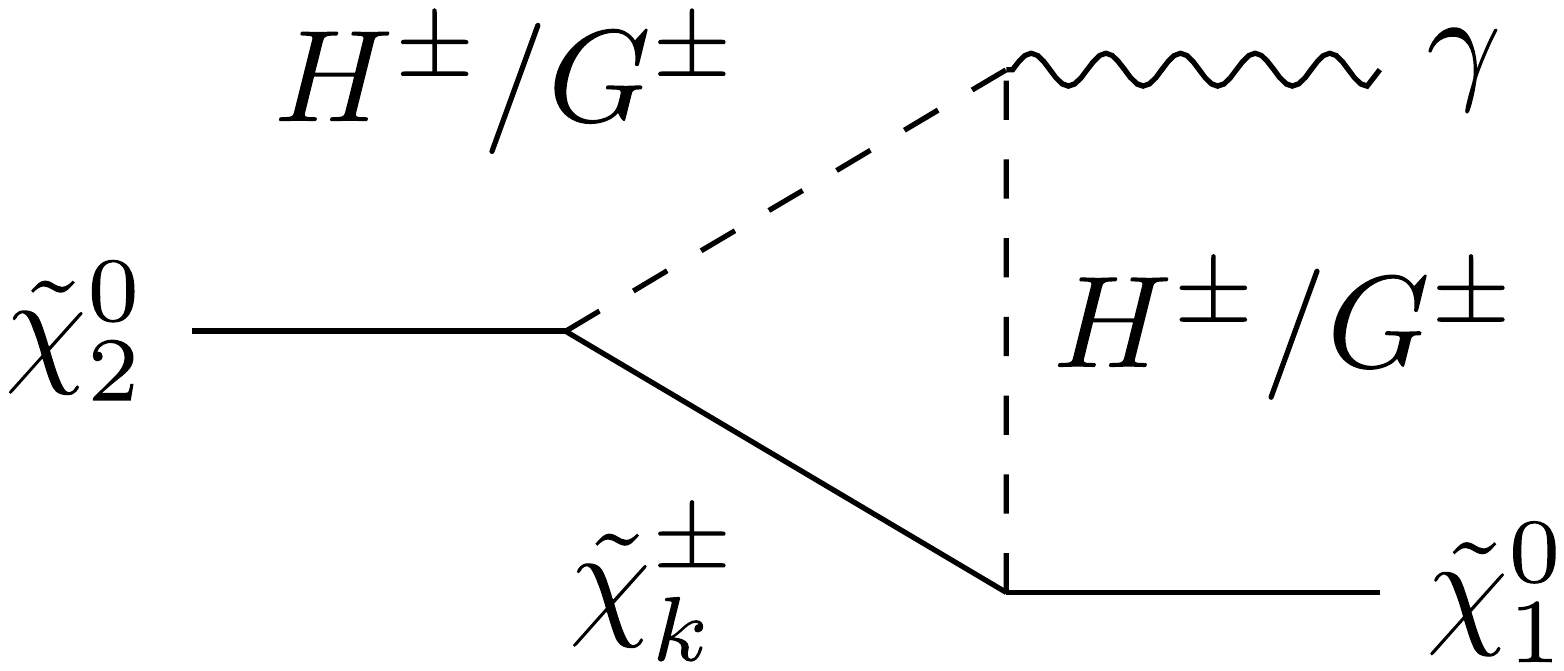}
        \end{subfigure}
        \caption{Triangle diagrams mediated by a charged Higgs ($H^\pm$) or Goldstone bosons ($G^\pm$) and a chargino ($\chak$) contributing to the $\chitwo$ radiative decay.}
    \label{fig:H_chi_loop}
    \end{figure}
    
    \begin{figure}
            \centering
            \begin{subfigure}{0.3\textwidth}
            \includegraphics[width=\textwidth]{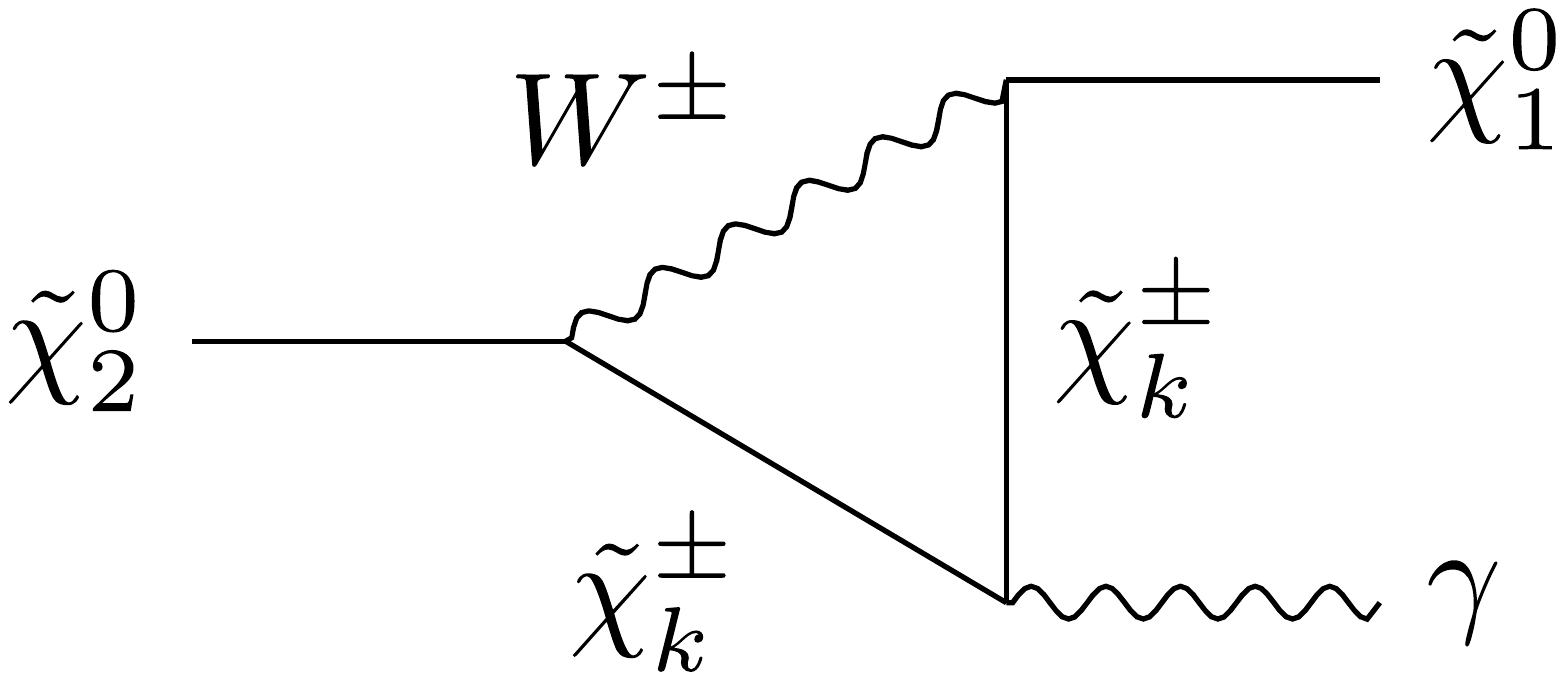}
            \end{subfigure}
            ~~~\begin{subfigure}{0.3\textwidth}
            \includegraphics[width=\textwidth]{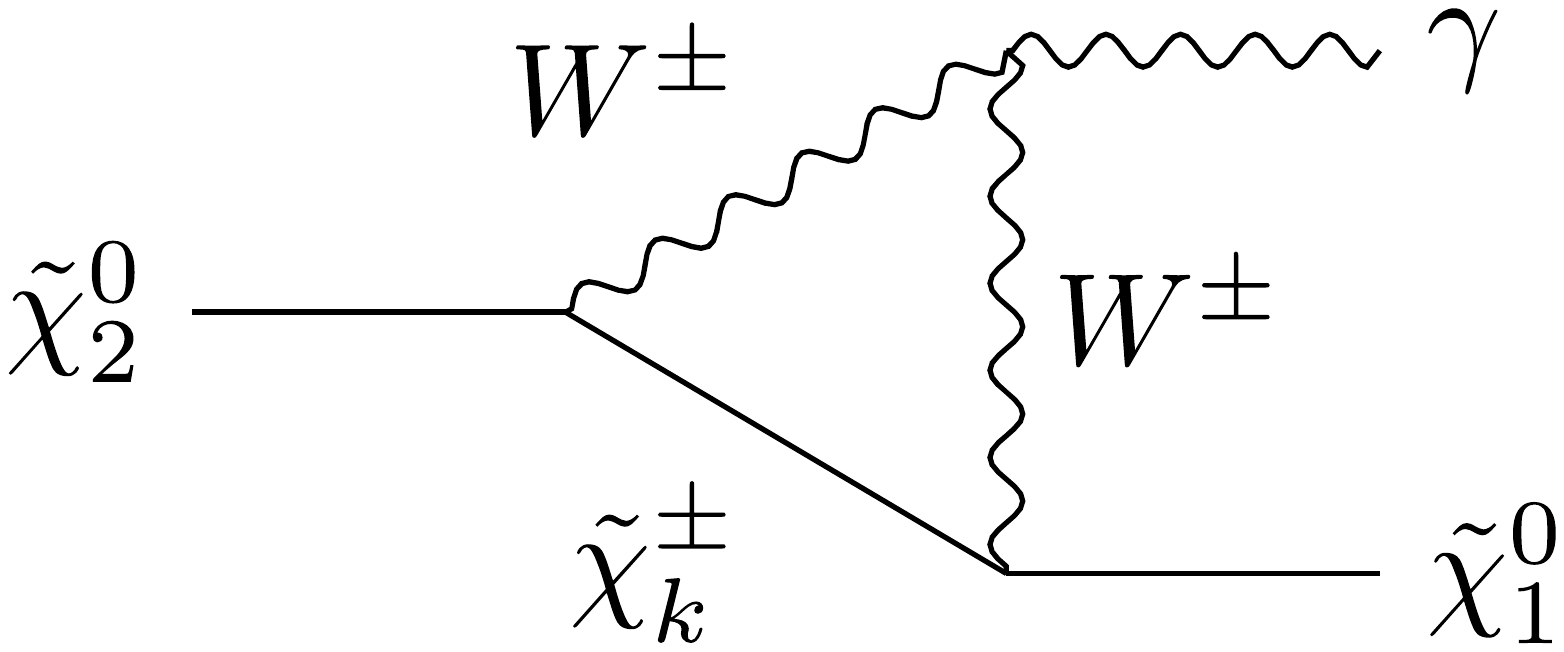}
            \end{subfigure}
            \caption{Triangle diagram mediated by a $W^\pm$-boson and a chargino ($\chak$) contributing to the $\chitwo$ radiative decay.}
            \label{fig:W_chi_loop}
    \end{figure}
    
    \item \bm{$W^{\pm}-\chak$} \textbf{triangle:} The corresponding diagrams are shown in \fig{fig:W_chi_loop}. The effective coupling generated via these diagrams is given by
    \begin{align} 
        g_{W^{\pm}/\tilde{\chi}^{\pm}} = \frac{-e g^2  m_{\chitwo}}{8 \pi^2}  \sum_k & \left\{ \left(G_L F_L-G_R F_R \right) \left[ (\xi_1\times \xi_2) m_{\chitwo} \left(I_2-J-K\right) + m_{\chione} \left(J-K\right) \right] \right. \nonumber \\
        & \left. \quad + 2\xi_1 m_{\chak} \left(G_L F_R-G_R F_L \right) J \right\} \;.
        \label{eq:g_W_chi}
    \end{align}
    
    The combinations of the couplings $(G_L F_L-G_R F_R)$ and $(G_L F_R-G_R F_L)$ are 
    \begin{align}
        G_L F_L-G_R F_R &= \mathbf{N}_{12} \mathbf{N}_{22} \left( \mathbf{V}_{k1}^2 - \mathbf{U}_{k1}^2 \right) + \frac{1}{2} \left( \mathbf{N}_{14} \mathbf{N}_{24} \mathbf{V}_{k2}^2 - \mathbf{N}_{13} \mathbf{N}_{23} \mathbf{U}_{k2}^2 \right) \\
        & \quad -\sqrt{\frac{1}{2}} \mathbf{V}_{k1} \mathbf{V}_{k2} \left( \mathbf{N}_{12} \mathbf{N}_{24} + \mathbf{N}_{14} \mathbf{N}_{22} \right) - \sqrt{\frac{1}{2}} \mathbf{U}_{k1} \mathbf{U}_{k2} \left( \mathbf{N}_{12} \mathbf{N}_{23} + \mathbf{N}_{13} \mathbf{N}_{22} \right) \;, \nonumber
    \end{align}
    \begin{align}
        \hspace*{-0.7cm}G_L F_R-G_R F_L &= \frac{1}{2} \mathbf{U}_{k2} \mathbf{V}_{k2} \left( \mathbf{N}_{13} \mathbf{N}_{24} - \mathbf{N}_{14} \mathbf{N}_{23} \right) + \sqrt{\frac{1}{2}} \mathbf{U}_{k1} \mathbf{V}_{k2} \left( \mathbf{N}_{12} \mathbf{N}_{24} - \mathbf{N}_{14} \mathbf{N}_{22} \right) \nonumber \\
        & \quad + \sqrt{\frac{1}{2}} \mathbf{U}_{k2} \mathbf{V}_{k1} \left( \mathbf{N}_{12} \mathbf{N}_{23} - \mathbf{N}_{13} \mathbf{N}_{22} \right) \;,
    \end{align}
    where $\mathbf{N}_{ij}$ and $\mathbf{U}_{ij}(\mathbf{V}_{ij})$ are the neutralino and chargino mixing matrix elements defined in \sect{sec:spec}. In the parameter space of interest, since the mixings in both the neutralino and the chargino sectors are highly suppressed, so are the factors $(G_L F_L-G_R F_R)$ and $(G_L F_R-G_R F_L)$. Therefore, the radiative decay mediated by the  $W^{\pm}-\tilde{\chi}^{\pm}$ triangle is highly suppressed.

    \end{enumerate}

\begin{figure}
    \includegraphics[width=\linewidth]{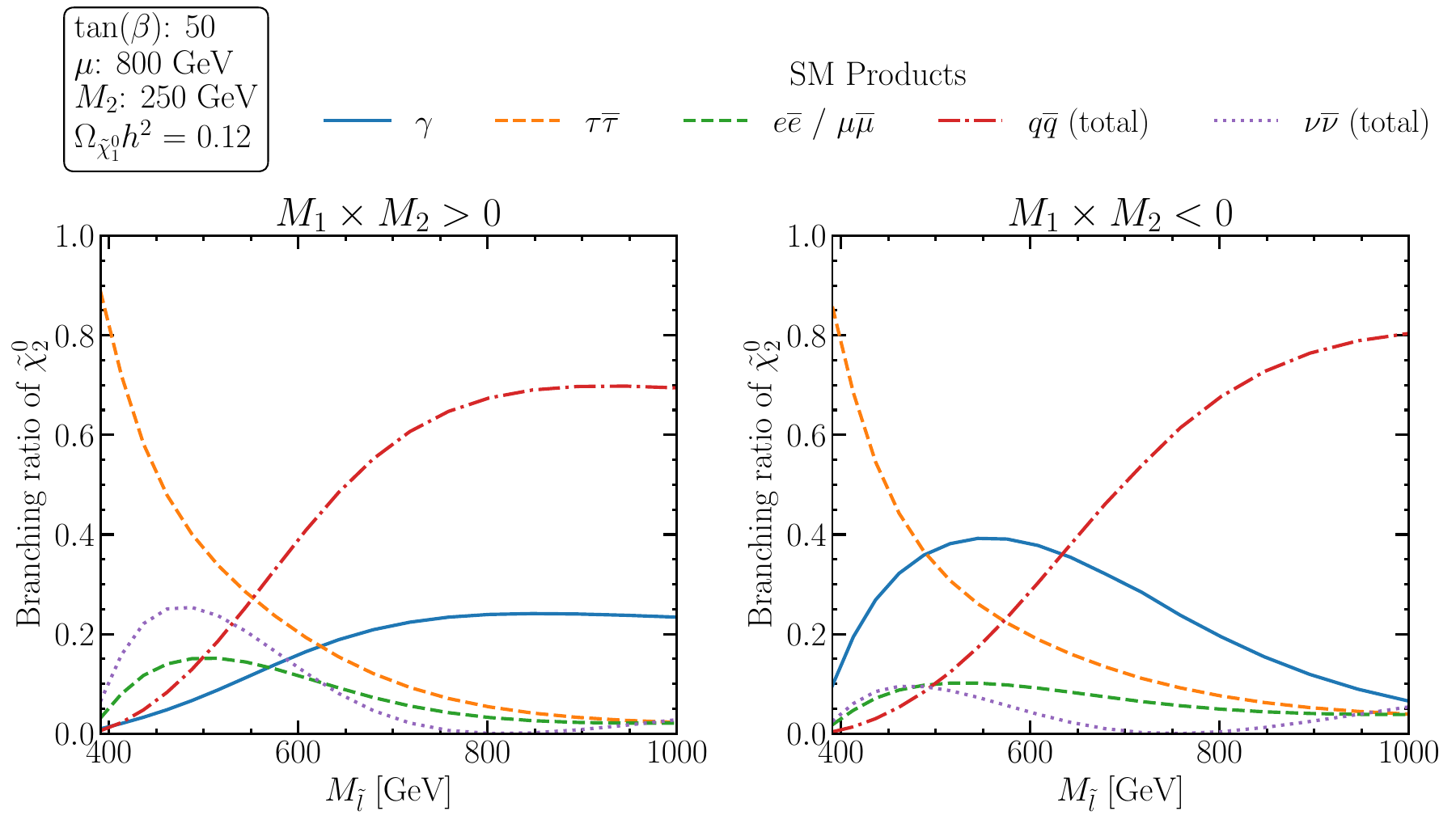}
    \caption{Plot of the branching ratios of the $\chitwo$ decay modes for a range of soft slepton masses, where $M_\slepton$ denotes the common soft breaking mass. All decay modes also contain a $\chione$ as a final product. Note that the $q\overline q$ ($\nu \overline \nu$) decay mode is the total branching ratio to all types of quarks (neutrinos), and the electron/muon branching ratio is the sum of the two. Here $M_1$ is fixed such that the neutralino relic density saturates the cosmological dark matter density, which sets $M_1 = 245$\,GeV with a variation of $\sim$2\,GeV across the slepton mass axis. This induces a mass splitting of $(\massplit) \sim 15-20$\,GeV. Note that $M_2 = 250$\,GeV is already excluded by LHC searches for $(M_1 \times M_2) > 0$, but there remains unconstrained parameter space for $(M_1 \times M_2) < 0$. For $(M_1 \times M_2) < 0$ and this choice of $\mu$ and $\tan(\beta)$, the radiative decay peaks between roughly $M_{\tilde l} \sim 450-700$\,GeV.}
    \label{fig:BRsOverSleptonMass}
\end{figure}

In summary, the dominant radiative decay channel is the one mediated by the $\sfermion-f$ triangle, which is enhanced for $(\xi_1 \times \xi_2)=-1$, corresponding to $(M_1\times M_2)<0$. 

To validate our analytical understanding, in \fig{fig:BRsOverSleptonMass}, we show the $\chitwo$ decay branching ratios obtained with \verb|SUSY-HIT|~\cite{Djouadi:2006bz}. For fixed benchmark values of $\tan\beta = 50$, $\mu = 800\,$GeV and $M_2 = 250\,$GeV, we show how the branching ratios change with the (generation-universal) soft slepton mass parameters $M_\slepton \equiv M_L = M_R$. Throughout these plots, we adjust $M_1$ such that we obtain a $\chione$ relic density of $\Omega_{\chione} h^2 = 0.12$, corresponding to a mass gap of the Bino-like neutralino and Wino-like next-to-lightest neutralino of $(\massplit) \sim20\,$GeV. The left panel is for the $(M_1 \times M_2) > 0$ case, while in the right panel, $(M_1 \times M_2) < 0$. In both cases, we observe that for small slepton masses, ($\chitwo \to \chione~\tau^+ \tau^-$) is the dominant decay mode. The tau-Yukawa is the largest of the charged leptons, and thus staus have the largest left-right mixing of the sleptons making the lighter stau the lightest slepton. The Higgs-mediated channel (\fig{fig:tree_h}) is enhanced by the large Yukawa coupling, and the sfermion-mediated channel (\fig{fig:tree_sf}) is enhanced by the light stau. As the slepton masses increase, slepton-mediated decays to leptons are suppressed, while the decays to quarks, which are mainly mediated by $Z/h$ and independent of slepton masses, become more and more important. The radiative decay, on the other hand, depends on the sign of $(M_1\times M_2)$. At lower slepton masses, the radiative decay branching ratio for $(M_1\times M_2)<0$ is significantly larger than in the $(M_1\times M_2)>0$ case, verifying the sign-dependence discussed above. 

For the particular choices of parameters in \fig{fig:BRsOverSleptonMass}, the radiative decay becomes the dominant decay mode (with branching ratios as larger as 40\,\%) in the range of the slepton masses $\sim 500-650$\,GeV for $(M_1\times M_2)<0$. The radiative decay becomes subdominant to the decays to quarks at heavier slepton masses due to the suppression of the $\sfermion-f$ triangles (\fig{fig:sf_f_loop}). Note that even in the case of $(M_1\times M_2)>0$, the branching ratio to radiative decay can reach $\mathcal{O}(10\,\%)$ and is larger than generally expected from the loop suppression; this is because of the kinematic suppression in the compressed region discussed at the beginning of this subsection. 

\section{Numerical Analysis}
\label{sec:Computations}

There is rich interplay between different phenomenological aspects in the MSSM, as we emphasized in the introduction and discussed in some detail in the preceding section. In particular, the DM phenomenology, contributions to the magnetic dipole moment of the muon, and decay patterns of the electroweakinos relevant for collider searches are highly interconnected. In order to gain a more quantitative understanding of the parameter space, we present results from a numerical study in this section. In particular, we are interested in identifying regions of parameter space where the MSSM features a viable DM candidate that explains the observed DM relic density while being compatible with all existing DM constraints without exacerbating electroweak fine-tuning problems. In addition, the MSSM parameter region should simultaneously explain the measured value of the magnetic dipole moment of the muon and satisfy constraints from searches for new particles at colliders. As anticipated in the preceding section, radiative decays of the Wino-like second-lighest neutralino can have large branching ratios, leading us to propose a new search channel at the LHC to better cover this interesting region of parameter space.

\begin{figure}
    \includegraphics[width=\linewidth, trim={0.5cm 0.5cm 2.5cm 1cm}, clip]{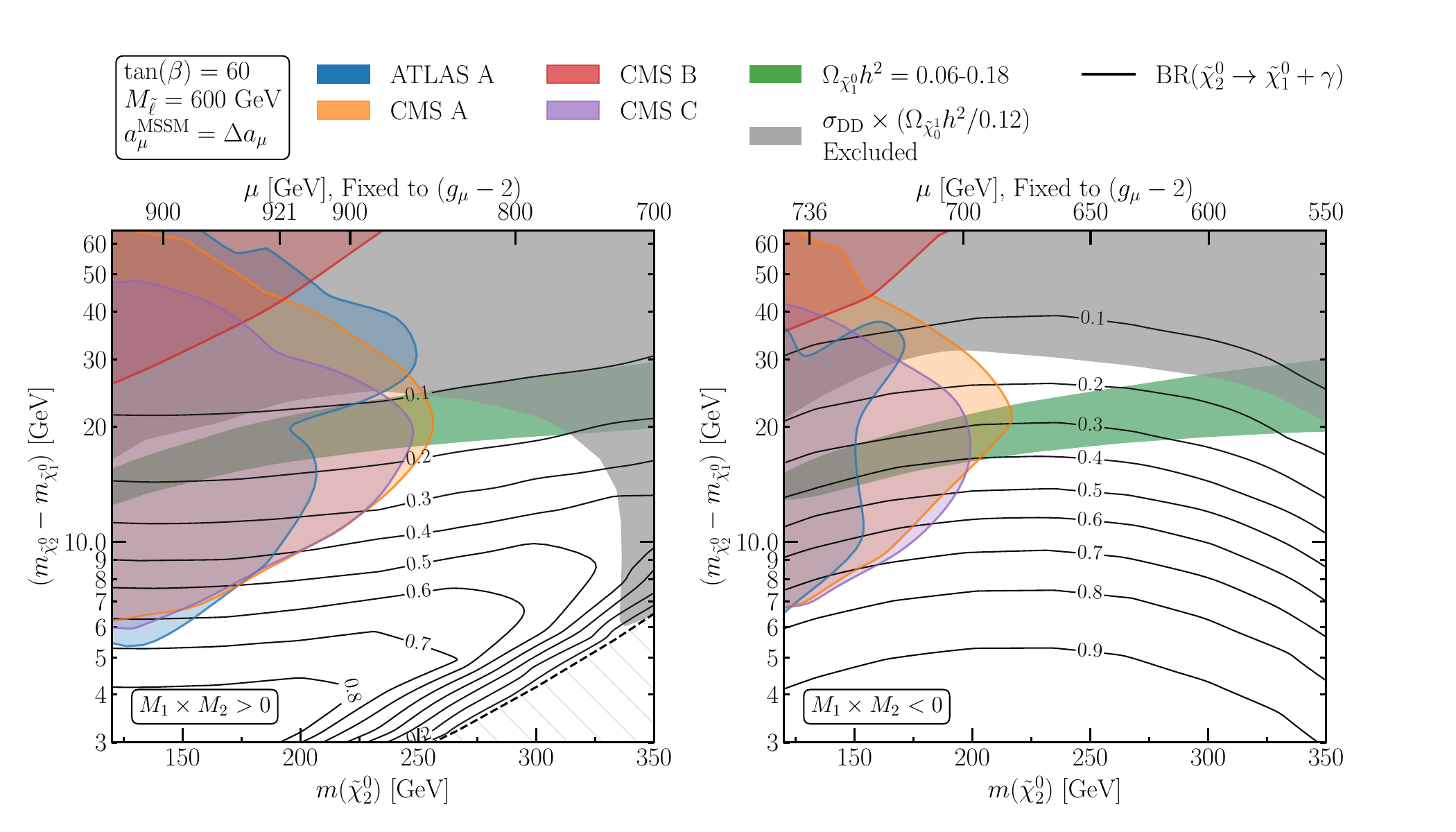}
    \caption{Projection of the MSSM parameter space onto the $m_{\chitwo}$ vs ($\massplit$) plane for $\tan\beta = 60$ and $M_{\slepton} = 600\,$GeV. For each point in the scan, we adjust the value of $\mu$, shown in the upper $x$-axis, such that the MSSM contribution to the magnetic dipole moment of the muon reproduces
    $\Delta a_\mu$. The left (right) panel is for $M_1 \times M_2 > 0$ $(M_1 \times M_2 < 0$). In both panels, the differently-colored regions in the upper left corner show the regions of parameter space ruled out by different LHC multilepton searches, as labeled in the legend, ``ATLAS~A''~\cite{ATLAS:2019lng}, ``CMS~A''~\cite{CMS-PAS-SUS-16-025}, ``CMS~B''~\cite{CMS:2017moi}, ``CMS~C''~\cite{CMS:2018kag}. The green band shows the region of parameter space where the $\chione$ relic density (approximately) matches the observed DM relic density ($\Omega_{\rm DM} h^2 = 0.12$); note that in the region above the green band neutralinos would overclose the Universe, while below the green band, $\Omega_{\chione} h^2 < \Omega_{\rm DM} h^2$, such that neutralinos would only be a subcomponent of DM. The gray-shaded region in the upper right corner is ruled out by the null results from direct detection experiments. The black lines are isocontours of the ``radiative decay'' branching ratio, ${\rm BR}(\chitwo \to \chione + \gamma)$. Finally, in the left panel, where ($M_1 \times M_2) > 0$, the hatched region in the lower right corner indicates ($\massplit$) values that cannot be realized due to level-repulsion in the neutralino mixing. For ($M_1 \times M_2) < 0$ (right panel), there is a region where the MSSM explains the observed DM relic density, the observed value of $a_\mu$, and is compatible with the null results from direct detection experiments and collider searches for $200\,{\rm GeV} \lesssim m_{\chitwo} \lesssim 325\,$GeV and with BR$ (\chitwo \to \chione + \gamma) \approx 0.2-0.4$.
    }
    \label{fig:exclusion_60_600}
\end{figure}

The results presented in this section are obtained with the following chain of numerical tools: first, we use \texttt{SUSY-HIT}~\cite{Djouadi:2006bz} to calculate the particle spectrum (at the one-loop level) and their decay rates. Note that as inputs for \texttt{SUSY-HIT} we use MSSM parameters defined at a scale given by the geometric mean of the two stop masses, except for $\tan\beta$ which is defined at the $Z$-boson pole. Note also that \texttt{SUSY-HIT} uses a calculation based on Refs.~\cite{Baer:2002kv, Haber:1988px} for the radiative decay widths of the neutralinos, which play a prominent role in this work. Second, we use \texttt{MicrOMEGAS~3.2}~\cite{Belanger:2018ccd} to calculate, to the leading order, the relic density, direct detection cross sections, and the value of the muon's anomalous magnetic moment. We note that the cross-sections for the co-annihilation processes controlling the relic abundance of $\chione$ are subject to higher-order corrections. Note that in our case, the most relevant co-annihilation processes involve only color-neutral particles, hence, higher-order corrections are controlled by the size of the electroweak couplings. Furthermore, while non-perturbative contributions such as the Sommerfeld effect (see, e.g., Ref.~\cite{Hryczuk:2011tq}) can be important for TeV-scale DM candidates, here we focus on Bino-like DM in the few hundred GeV mass range where such effects are small~\cite{Beneke:2022rjv}. Nonetheless, the corrections to the leading-order co-annihilation cross section from higher-order and non-perturbative corrections can amount to changing the mass difference between the Wino-like and the Bino-like states leading to the correct relic density by a few GeV; we account for these corrections by conservatively assuming an uncertainty of ${\sim}\,50\,\%$ for the relic density, see, e.g., the green band in \fig{fig:exclusion_60_600}. We use our own code to compare the direct detection cross sections to the null results from the LUX-ZEPLIN (LZ)~\cite{LZ:2022ufs} and PICO-60~\cite{PICO:2017tgi} experiments; these experiments have published the currently leading limits on the spin-independent WIMP-nucleon as well as spin-dependent WIMP-neutron and WIMP-proton cross sections. When comparing the MSSM cross sections against these experimental results, we re-scale the reported upper limit by the predicted neutralino relic density $\Omega_{\chione}$ as $0.12/(\Omega_{\chione} h^2)$, implicitly assuming that the local neutralino density scales linearly with the average cosmological relic density. In the same fashion, we also compare the direct detection cross sections to the projected reach of the full exposure of the XENONnT~\cite{XENON:2020kmp}, LZ~\cite{LZ:2018qzl}, and PICO-40L~\cite{Krauss:2020ofg} experiments. Third, we use \texttt{CheckMATE~2.2}~\cite{Dercks:2016npn, deFavereau:2013fsa, Cacciari:2011ma, Cacciari:2008gp, Read:2002hq} to re-cast the null-results from LHC searches for new particles. For searches for electroweakinos and sleptons, re-interpreting LHC bounds is a non-trivial exercise since multiple production channels and decay modes can contribute to a given final state, affecting the kinematics. Specifically, we generate Monte Carlo events for the productions of charginos/neutralinos and sleptons via proton-proton collisions at the LHC with \texttt{Madgraph~3.2}~\cite{Alwall:2014hca}, implement hadronization, showering, and $\chitwo$ and $\chaone$ decays with \texttt{Pythia~8.2}~\cite{Sjostrand:2014zea}, and simulate detector effects with \texttt{Delphes~3}~\cite{deFavereau:2013fsa}. 
The resulting final-state distributions are then directly compared to the results from a library of LHC searches; the version of \texttt{CheckMATE} we employ uses 39 searches at $\sqrt{s} = 13\,$TeV. As we will see, most important in the compressed region (where $\massplit \sim 10 - 30\,$GeV) are searches in the dilepton final states by ATLAS~\cite{ATLAS:2019lng} which made use of $\mathcal{L} = 139\,$fb$^{-1}$ of data and CMS with $\mathcal{L} = 12.9\,$fb$^{-1}$~\cite{CMS-PAS-SUS-16-025} and $\mathcal{L} = 35.9\,$fb$^{-1}$~\cite{CMS:2018kag} of data. We note that CMS searches with the full 139\,fb$^{-1}$ LHC~Run~2 data set would likely strengthen the CMS bounds by up to $\mathcal{O}(10)\,$GeV in the electroweakino masses (for example, compare Refs.~\cite{CMS:2018kag, CMS:2021edw}). These searches are, however, not available in \texttt{CheckMATE}.

In all results presented in this work, we set the (generation-universal) soft squark mass parameters as well as the gluino mass parameter to $M_{\widetilde{Q}_L} = M_{\widetilde{u}_R} = M_{\widetilde{d}_R} = M_3 = 2.5\,$TeV. By setting the $A$-term of the up-type squarks to $A_t = 3.5\,$TeV, we achieve appropriate radiative corrections to the SM-like Higgs boson mass to obtain $m_h \approx 125\,$GeV throughout the region of parameter space we explore. We set the $A$-terms for the bottom-type squarks and the sleptons to $A_b = -100\,$GeV and $A_\tau = -250\,$GeV, respectively. Finally we set the mass parameter for the non-SM-like heavy Higgs bosons to $M_A = 2.5\,$TeV.
We present results in two different two-dimensional projections of the remaining MSSM parameter space to be discussed below. For both projections, we fix the values of $\tan\beta$ and the generation-universal soft slepton mass parameters $M_{\slepton} \equiv M_L = M_R$ to benchmark values in the range $50 \leq \tan\beta \leq 70$ and $550\,{\rm GeV} \leq M_{\slepton} \leq 700\,$GeV. The slepton masses are bounded from below by slepton production searches in the LHC, see Ref. \cite{ATLAS:1908.08215}. By choosing $M_\slepton = 600$\,GeV, these bounds can be avoided due to the large dependence of the slepton production cross section on $M_\slepton$. Furthermore, we choose $M_2$ and $\mu$ to take positive values, while $M_1$ can take either sign.

For the first projection, we show results projected onto the plane spanned by the mass of the Wino-like second-lightest neutralino ($m_{\chitwo}$) and the mass difference between the Bino-like lightest neutralino and the second-lightest neutralino, ($\massplit$). Note that this plane is often chosen by the LHC collaborations to present results from searches for electroweakinos aimed at the compressed region. 
We scan over the Bino and Wino mass parameters $M_1$ and $M_2$ and for each point of the scan, we adjust the value of the Higgsino mass parameter $\mu$ such that the MSSM contribution to the magnetic dipole moment of the muon accounts for the difference between the measured value of $a_\mu$~\cite{Muong-2:2021ojo} and its SM prediction~\cite{Aoyama:2020ynm}.

In \fig{fig:exclusion_60_600}, we fixed $\tan\beta = 60$ and $M_\slepton = 600\,$GeV, while the corresponding values of $\mu$ are shown by the ticks on the upper edge of both panels. Note that these values of $\mu$ are not a monotonic function of $m_\chitwo$. Also observe that since $a_\mu^{\rm MSSM}$ depends on $M_1$, there is a variation in the value of $\mu$ for which $a_\mu^{\rm MSSM} = \Delta a_\mu$ with $(\massplit)$; this variation is of order 5\% across the range of values of $(\massplit)$ shown in \fig{fig:exclusion_60_600}.

In the left panel of \fig{fig:exclusion_60_600}, we show results for $(M_1 \times M_2) > 0$, while in the right panel, we show results for $(M_1 \times M_2) < 0$. To start, we can note that across the range of $m_\chitwo$ values shown in \fig{fig:exclusion_60_600}, $120\,{\rm GeV} \lesssim m_\chitwo \lesssim 350\,$GeV, we find that the lightest neutralino has the right relic density to account for the DM for $(\massplit) \sim 10 - 30\,$GeV. For smaller $(\massplit)$, Bino-Wino co-annihilation becomes so efficient that the $\chione$'s would only comprise a subcomponent of DM, while for larger $(\massplit)$, co-annihilation is no longer sufficiently efficient to avoid the $\chione$'s overclosing the universe. Next, let us discuss the behavior of the constraints from direct detection experiments shown by the gray-shaded region in \fig{fig:exclusion_60_600}. Focusing on the green band where neutralinos make up all of the DM, we can see that for both signs of $(M_1 \times M_2)$, the null results from direct detection experiments rule out the region of parameter space where $m_\chitwo \gtrsim 300\,$GeV. However, we can also note that the values of $\mu$ required for $a_\mu^{\rm MSSM} = \Delta a_\mu$ are roughly 200\,GeV {\it smaller} for $(M_1 \times M_2) < 0$ than those for $(M_1 \times M_2) > 0$.\footnote{For $(M_1\times M_2) > 0$, there is another branch with yet larger $\mu$-values satisfying $\Delta a_\mu$ and direct detection bounds, which we do not consider in this analysis. See, for example, Refs.~\cite{Abdughani:2019wai,Chakraborti:2020vjp,Chakraborti:2021kkr,Chakraborti:2021dli,Chakraborti:2021mbr} for studies of this ``large-$|\mu|$'' region of parameter space.} Thus, the fact that the regions ruled out by direct detection constraints visually appear similar for both signs in the projection of \fig{fig:exclusion_60_600} reflects the effectiveness of the suppression of the direct detection cross section for $(M_1 \times \mu) < 0$ discussed in \sect{sec:relicDensity}. Recall that we choose $M_A=2.5$\,TeV for all numerical results in this work, and choose $\tan\beta=60$ for \fig{fig:exclusion_60_600}. Note that the current LHC bounds rule out $M_A\lesssim 2$\,TeV for $\tan\beta=60$ \cite{ATLAS:2020zms,CMS:2022goy}. For $(M_1\times M_2)>0$, smaller $M_A$ leads to larger spin-independent direct detection cross section; while for $(M_1\times M_2)<0$, smaller $M_A$ leads to smaller direct detection cross sections. Hence for $M_A=2$\,TeV, the difference between the regions of parameter space disfavored by direct detection bounds would be more pronounced between the different signs of $(M_1\times M_2)$ than for $M_A=2.5$\,TeV as shown in \fig{fig:exclusion_60_600}.

In \fig{fig:exclusion_60_600} we also show regions of parameter space ruled out by the null results from searches at the LHC by the various colored regions in the upper left corner of each panel. The most relevant current searches at the LHC for this region of parameter space are searches for events with multiple charged leptons with relatively low transverse momenta. Comparing between the $(M_1 \times M_2) > 0$ case (left) and the $(M_1 \times M_2) < 0$ case (right), we can see that the null-results from these collider searches rule out a significantly smaller region of the parameter space for $(M_1 \times M_2) < 0$ than for $(M_1 \times M_2) > 0$. At the same time, we can read off that, as anticipated in \sect{sec:decay}, radiative decays of the second-lightest neutralino have much larger branching ratios for $(M_1 \times M_2) < 0$ than for $(M_1 \times M_2) > 0$. The larger the radiative decay branching ratio is, the smaller the branching ratios of $\chitwo$ into final states including charged leptons which the multi-lepton searches are sensitive to. 

As shown in \fig{fig:exclusion_60_600}, combining the direct detection constraints and the collider bounds, we find that for $(M_1 \times M_2) > 0$ most of the parameter space where neutralinos account for all of the DM is ruled out. For ($M_1 \times M_2) < 0$, instead, there is a region where the MSSM explains the observed DM relic density, the observed value of $a_\mu$, and is compatible with the null results from direct detection experiments and collider searches for $200\,{\rm GeV} \lesssim m_{\chitwo} \lesssim 325\,$GeV; note that in this region of parameter space, the radiative decay ($\chitwo \to \chione + \gamma$) has branching ratios of $20-40\,\%$. Recall that in \fig{fig:exclusion_60_600}, we chose $\tan\beta = 60$ and $M_\slepton = 600\,$GeV. Focusing on the interesting case $(M_1 \times M_2) < 0$, we show in \app{app:Appendix} results equivalent to the right panel of \fig{fig:exclusion_60_600} for a range of values of $\tan\beta = 50-70$ and $M_\slepton = 550-700\,$GeV. These results feature similar qualitative behavior, but illustrate that the specific boundaries of the allowed parameter space as well as the values of the radiative decay branching ratios depend on $M_\slepton$ and $\tan\beta$. 

\begin{figure}
    \includegraphics[width=\linewidth]{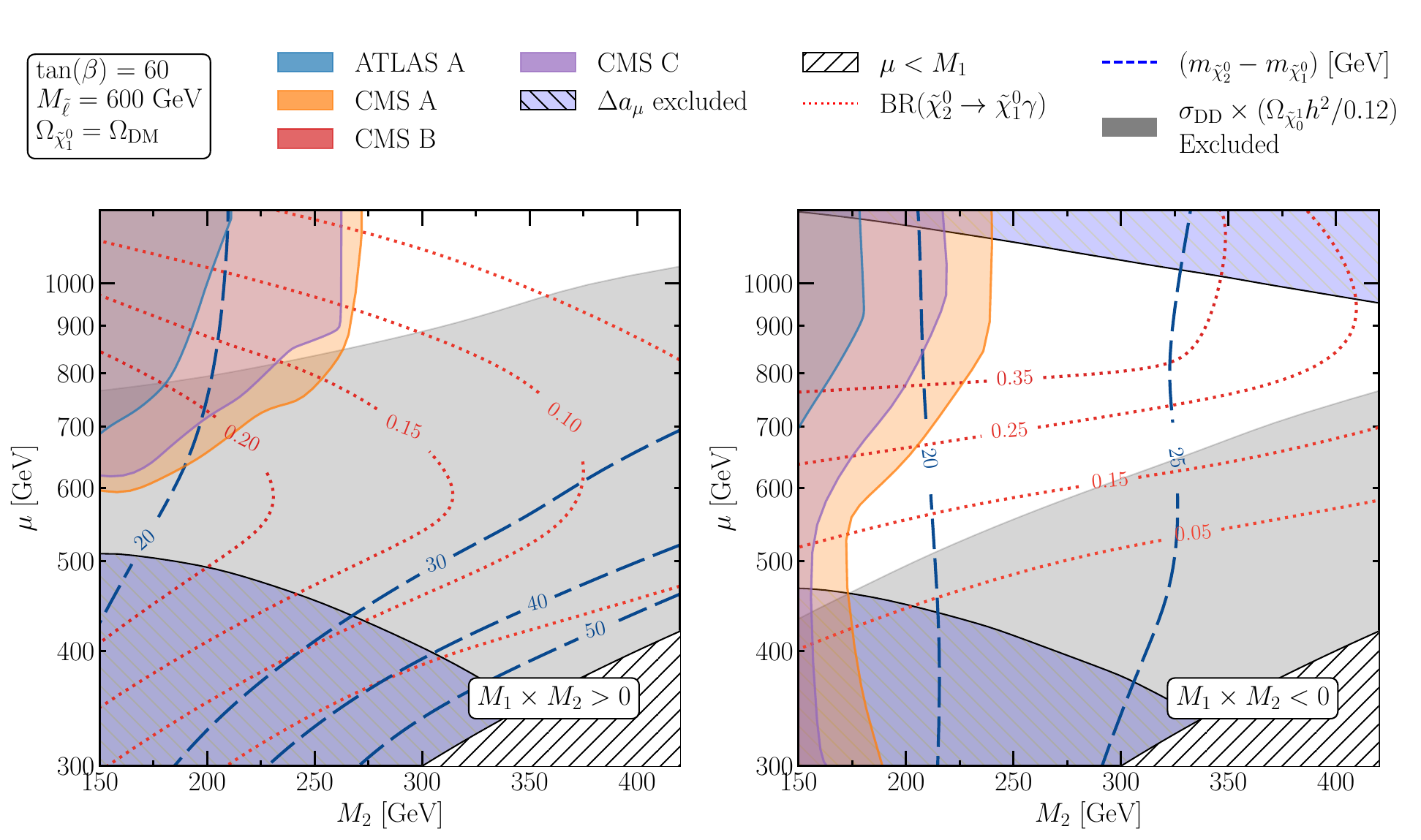}
    \caption{Projection of the MSSM parameter space onto the $M_2$ vs $\mu$ plane for $\tan\beta = 60$ and $M_\slepton = 600\,$GeV. In this projection, we adjust $M_1$ at each point of the scan such that the neutralinos make up all of the DM, $\Omega_\chione = \Omega_{\rm DM}$. As in \fig{fig:exclusion_60_600}, the left panel is for $(M_1 \times M_2) > 0$ and the right panel is for $(M_1 \times M_2) < 0$, regions of parameter space ruled out by collider searches are shown by the differently-colored regions in the upper left corners, and the constraints from direct detection experiments are shown by the gray shaded area. The light-blue shaded region is where the MSSM contribution to the muon's magnetic moment deviates more than 2$\sigma$ from the observed value of $\Delta a_\mu$. In the lower-right hatched region, $\mu$ becomes comparable to $M_1$ which drastically changes the phenomenology. The dashed blue contours show the mass splitting ($\massplit$) in units of GeV (which we adjust to reproduce the observed relic density), and the dotted red contours show the radiative decay branching ratio of $\chitwo$. The white region is compatible with all current constraints; note that for $(M_1 \times M_2) < 0$, the allowed region of parameter space with $M_2 \sim 250-400\,$GeV features large values of the $\chitwo$ radiative decay branching ratio of $20-40\,\%$.}
    \label{fig:mu_m2_plane}
\end{figure}

In \fig{fig:mu_m2_plane} we show results in the $M_2$ vs $\mu$ plane. We fix $\tan\beta = 60$ and $M_\slepton = 600\,$GeV and scan over the Wino and Higgsino mass parameters $M_2$ and $\mu$. At each point of the scan, we adjust the Bino mass parameter $M_1$ such that the relic density of the (Bino-like) lightest neutralino ($\Omega_{\chione} h^2$) matches the observed DM relic density ($\Omega_{\rm DM} h^2 = 0.12$). Note that Bino-Wino co-annihilation becomes less efficient for larger ($\massplit$). Thus, adjusting $M_1$ such that $\Omega_\chione = \Omega_{\rm DM}$ corresponds to choosing the maximal value of ($\massplit$) allowed; for larger ($\massplit$), neutralinos would overclose the universe. Since the decay products from $\chitwo$ decays become more energetic for larger values of ($\massplit$), the region of the parameter space with maximal mass splitting is particularly interesting for collider searches. Compared to the first projection in the $m_\chitwo$ vs ($\massplit$) plane (see \fig{fig:exclusion_60_600}), some effects are easier to observe in \fig{fig:mu_m2_plane}, while others are more masked. We are again showing results for $(M_1 \times M_2) > 0$ in the left panel, while the right panel is for $(M_1 \times M_2) < 0$. 

Let us start by discussing the values of ($\massplit$) producing $\Omega_\chione = \Omega_{\rm DM}$, shown by the blue dashed contours in \fig{fig:mu_m2_plane}. For most of the shown region of parameter space, we have $(\massplit) \sim 20-30 (50) \,$GeV for $M_1 \times M_2 <0$ ($M_1\times M_2 >0$). We observe that, in the lower right region of each panel, the requirement of Wino-Bino co-annihilation is somewhat relaxed and ($\massplit$) can take larger values. This is because the Higgsino mass parameter $\mu$ becomes comparable to the mass of the Bino-like $\chione$, and the correct relic density can be achieved by sufficiently large Higgsino-Bino mixing (the well-tempered neutralino). However, this region of parameter space is entirely ruled out by direct detection constraints. 

Next, let us discuss the behavior of the direct detection constraints. As we can see, these constraints rule out the region of parameter space where the values of $\mu$ become too close to the mass of the lightest neutralino, since then the Higgsino admixtures to the lightest neutralino, which control the direct detection cross sections, become too large. Comparing the two panels in \fig{fig:mu_m2_plane}, we see that for $(M_1 \times M_2) < 0$, much smaller values of $\mu$ are allowed by the direct detection constraints than for the $(M_1 \times M_2) > 0$ case, clearly showing the effect of the partial cancellations of the various amplitudes contributing to the spin-independent direct detection cross section discussed in \sect{sec:relicDensity}. 

In \fig{fig:mu_m2_plane}, the regions where the MSSM contribution to the magnetic dipole moment of the muon ($a_\mu^{\rm MSSM}$) differ by more than $2\sigma$ from the observed value of $\Delta a_\mu = (25.1 \pm 5.9) \times 10^{-10}$ are shown by the light-blue shaded regions. Comparing the two panels, we can note that the region where $a_\mu^{\rm MSSM}$ is within $2\sigma$ of $\Delta a_\mu$, i.e., the region in between these light-blue shaded regions, is at values of $\mu$ roughly $100-200\,$GeV smaller for $(M_1 \times M_2) < 0$ than for $(M_1 \times M_2) > 0$. This visualizes the behavior discussed in \sect{sec:gm2}: recall that the leading contributions to $a_\mu^{\rm MSSM}$ stem from the Bino-smuon and the chargino-sneutrino loops. For $(M_1 \times M_2) > 0$, and $\mu$ taking the same sign as $M_2$ as we assume throughout, both contributions are positive. For $(M_1 \times M_2) < 0$, on the other hand, the Bino-smuon contribution to $a_\mu^{\rm MSSM}$ becomes negative, such that a smaller value of $\mu$ is required in order for the chargino-sneutrino contribution to become sufficiently large for the total $a_\mu^{\rm MSSM}$ to be able to explain the observed value $\Delta a_\mu$~\cite{Baum:2021qzx}.

Finally, in \fig{fig:mu_m2_plane} we show the regions of parameter space ruled out by LHC searches for electroweakinos and sleptons with the differently-color shaded regions visible in the upper left corners. As we already observed in the $m_\chitwo$ vs ($\massplit$) plane, these constraints are again weaker for $(M_1 \times M_2) < 0$ than for $(M_1 \times M_2) > 0$, which is related to the branching ratio of ($\chitwo \to \chione + \gamma$) taking much larger values for the former case, suppressing the reach of these multi-lepton searches. 

In summary, both in \fig{fig:exclusion_60_600} and in \fig{fig:mu_m2_plane}, we can observe that for $(M_1 \times M_2) < 0$, there is a sizable region of parameter space where the MSSM explains the observed DM relic density, the observed value of $a_\mu$, and is compatible with the null results from direct detection experiments and collider searches for $200\,{\rm GeV} \lesssim m_{\chitwo} \lesssim 350\,$GeV; note that in this region of parameter space, the radiative decay ($\chitwo \to \chione + \gamma$) has branching ratios of $20-40\,\%$. The current LHC searches do not probe this region of parameter space, in part because the large values of the $(\chitwo \to \chione + \gamma)$ branching ratio suppress the decays into the charged-lepton final states that the current searches mainly use. On the other hand, these large values of the $(\chitwo \to \chione + \gamma)$ branching ratio suggest a new detection channel at the LHC that could be used to better cover this region of the MSSM parameter space, as we discuss in the following section. Before moving on, let us comment on the reach of future direct detection experiments, in particular, the projected reach of XENONnT and LZ with their full envisaged exposure. Most of the region of parameter space for both $(M_1 \times M_2) > 0$ and $(M_1 \times M_2) < 0$ where the MSSM satisfies all of our requirements shown in Figs.~\ref{fig:exclusion_60_600},~\ref{fig:mu_m2_plane}, and~\ref{fig:more_plots} is within the reach of these experiments. Thus, future DM direct detection experiments might make a discovery in this well-motivated region of parameter space. However, note that for our numerical results, the spin-independent direct detection cross sections are suppressed for $(M_1 \times M_2) < 0$ due to the partial cancellations of the various contributing amplitudes, and this cross section would be further suppressed if the values of $M_1$, $\mu$, $\tan\beta$, and $M_A$ are tuned to realize the ``generalized blind-spot'' solutions~\cite{Huang:2014xua} (see the discussion in \sect{sec:relicDensity}). Such solutions can suppress the spin-independent direct detection cross sections to very small values at the cost of tuning of the parameters. In this case, probes of the spin-dependent direct detection cross sections will become a powerful probe (see, for example, Refs.~\cite{Cohen:2011ec,Baum:2017enm} for related discussions).

\section{A new search channel at the LHC}
\label{sec:LHC}
As discussed in the previous section, there is a region of MSSM parameter space, where the neutralino spectrum is compressed ($\massplit \sim 10 - 30$\,GeV) and $(M_1 \times M_2) <0$, that is of special phenomenological interest. As we have seen, this region of parameter space is beyond the reach of current LHC compressed region searches, see, e.g., Refs.~\cite{ATLAS:2018ojr,ATLAS:2019lff,CMS:2020bfa,ATLAS:2021moa,CMS:2021cox,CMS:2021few,ATLAS:2021yqv,ATLAS:2022zwa,CMS:2022sfi,CMS:2019san}, since the decays of $\chitwo$ into final states containing charged leptons are suppressed by the large radiative decay branching ratio, ${\rm BR}(\chitwo \to \chione + \gamma) \sim 20-40\,\%$. Here, we perform a first study of the signal that could arise from these radiative decays: Wino-like charginos and neutralinos have sizable production cross sections at the LHC, in particular in the ($pp \to \chitwo + \chaone$) channel mediated by a $W^\pm$ in the $s$-channel. If the $\chitwo$ decays radiatively, the collider signature of this search is a relatively soft photon (with transverse momentum $\ptgamma \lesssim \massplit$) accompanied by the similarly soft visible decay products of the $\chaone$, see \fig{fig:ChannelFeyn} for an illustration. While some searches for photons and missing transverse energy (\MET) exist at the LHC, they are not well-suited to cover this region of parameter space. CMS has several analyses which search for a single hard photon and missing transverse energy (\MET) \cite{CMS:2018fon, CMS:2017brl, CMS:2017qca}, as does ATLAS \cite{ATLAS:2022ckd, ATLAS:2014kci}. To enhance the signature of the radiative decay channel with a soft photon and \MET at the LHC, one can consider events where the ($\chitwo + \chaone$) Wino pair is produced in conjunction with a hard initial state radiation (ISR) jet (j). While such an ISR boost will only lead to a $\mathcal{O}(1)$ increase of $\ptgamma$ due to the kinematics of the event, the recoil of the ($\chitwo + \chaone$) system against the ISR jet in ($pp \to \chitwo + \chaone + j$) events leads to a sizable increase in the missing transverse energy ($\MET$) of such events.

In order to analyze the kinematics of our signal events of interest, we simulate ($pp \to \chitwo + \chaone + j$) events at the $\sqrt{s} = 13\,$TeV LHC and analyze the ($j + \gamma + \MET + X$) final state arising from ($\chitwo \to \chione + \gamma$) radiative decays and chargino decays. We will focus on the distributions of $\ptgamma$ and $\MET$ in such events; the visible decay products (``$X$'') from the $\chaone \to \chione + X$ decays can in principle be used as additional handles to differentiate signal events arising from this channel from backgrounds. For our Monte Carlo event generation chain, we use \texttt{Madgraph~3.2}~\cite{Alwall:2014hca} to generate the hard ($pp \to \chitwo + \chaone + j$) events at leading order, \texttt{Pythia~8.2}~\cite{Sjostrand:2014zea} for hadronization, showering, and modeling of the $\chitwo$ and $\chaone$ decays, and \texttt{Delphes~3}~\cite{deFavereau:2013fsa} with the default ATLAS card to simulate the detector. We use \texttt{Prospino~2.1}~\cite{Beenakker:1996ed} to calculate the K-factor for next-to-leading order (NLO) corrections for electroweakino production, and we estimate that corrections would increase the LHC cross sections by about 20-30\% in our regions of parameter space, which we take into account by applying a common K-factor of 1.25. We perform a first event selection using the following set of cuts:
\begin{itemize}
    \item At the truth level (i.e. the level of the hard event), we require one (ISR) jet satisfying $p_T^j > 100\,$GeV and pseudorapidity $\eta^j < 5$. 
    \item At the detector level (i.e. after \texttt{Delphes}), we required events to contain one reconstructed photon with $\ptgamma > 10\,$GeV and $\eta^\gamma < 2.5$.
\end{itemize}
Additional cuts on $\ptgamma$, $\MET$, or any other collider observable can of course be added, and we will discuss some examples further below.

\begin{figure}
    \includegraphics[width=0.49\linewidth]{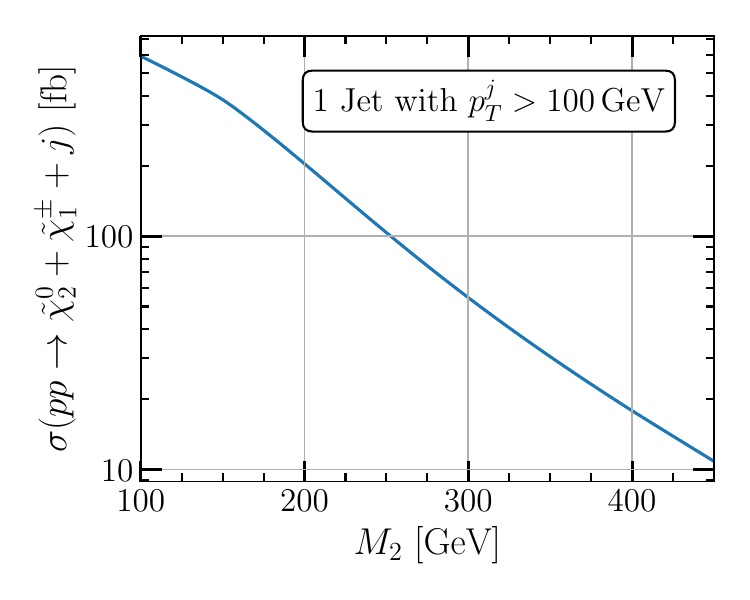}
    \includegraphics[width=0.49\linewidth]{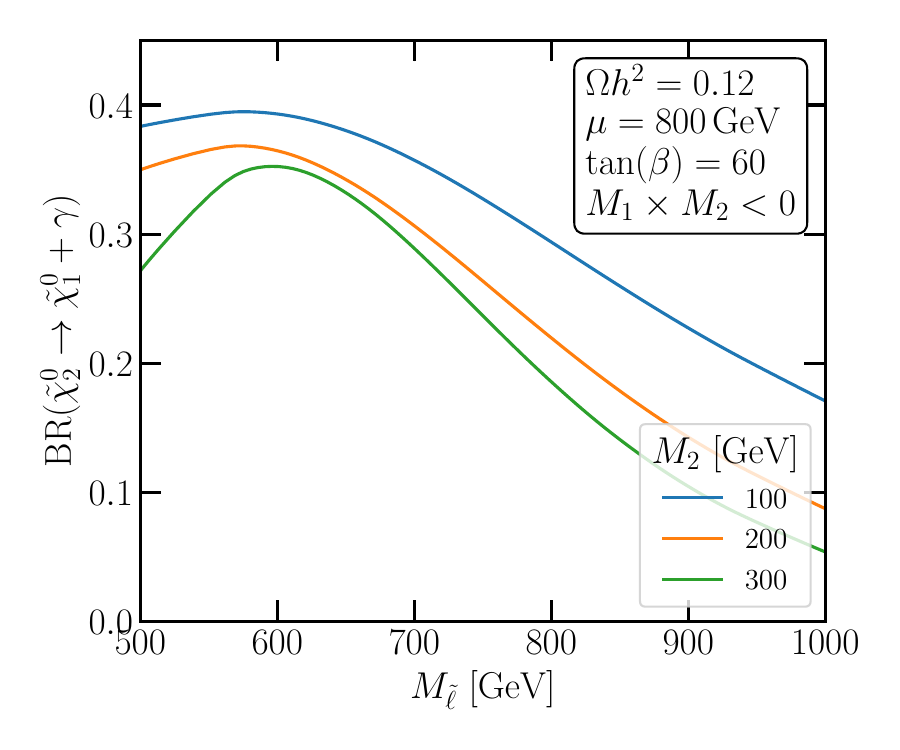}
    \caption{{\it Left:} Leading order production cross section $\sigma(pp \to \chitwo + \chaone + j)$ at the $\sqrt{s} = 13\,$TeV LHC, requiring a jet with $p_T^j > 100$\,GeV, and including an estimated 25\% NLO contribution. {\it Right:} Radiative decay branching ratio as a function of the slepton mass, for different values of the Wino mass parameter as denoted in the legend. The radiative decay branching ratio is largest for slepton masses from about 100\,GeV larger than the neutralino mass up to several hundreds of GeV.}
    \label{fig:xsec_BR}
\end{figure}

In the left panel of \fig{fig:xsec_BR}, we show $\sigma(pp \to \chitwo + \chaone + j)$ vs $M_2$, demanding $p_T^j > 100\,$GeV. In the most relevant range of parameter space, $M_2 \lesssim 350\,$GeV, we find $\sigma(pp \to \chitwo + \chaone + j) \gtrsim 20\,$fb, to be compared with the expected LHC Run~3 luminosity of $\mathcal{L} = 250\,{\rm fb}^{-1}$. The right panel of \fig{fig:xsec_BR} shows the radiative decay branching ratio, ${\rm BR}(\chitwo \to \chione + \gamma)$, as a function of the soft slepton mass parameter ($M_\slepton$) for various choices of the Wino mass parameter. As discussed in more detail in \sect{sec:decay}, \fig{fig:xsec_BR} reflects the suppression of ${\rm BR}(\chitwo \to \chione + \gamma)$ for light sleptons, where ($\chitwo \to \chione+l + \bar{l}$) three-body decays are prominent; and also the suppression for large slepton masses, where all processes mediated by sleptons are strongly suppressed and hadronic decays begin to dominate. Quite generally, for $\tan(\beta) \sim 60$ we find that ${\rm BR}(\chitwo \to \chione + \gamma)$ is largest for slepton mass parameters in the range $M_\slepton \sim 500-700\,$GeV.

\begin{figure}
    \includegraphics[width=0.49\linewidth]{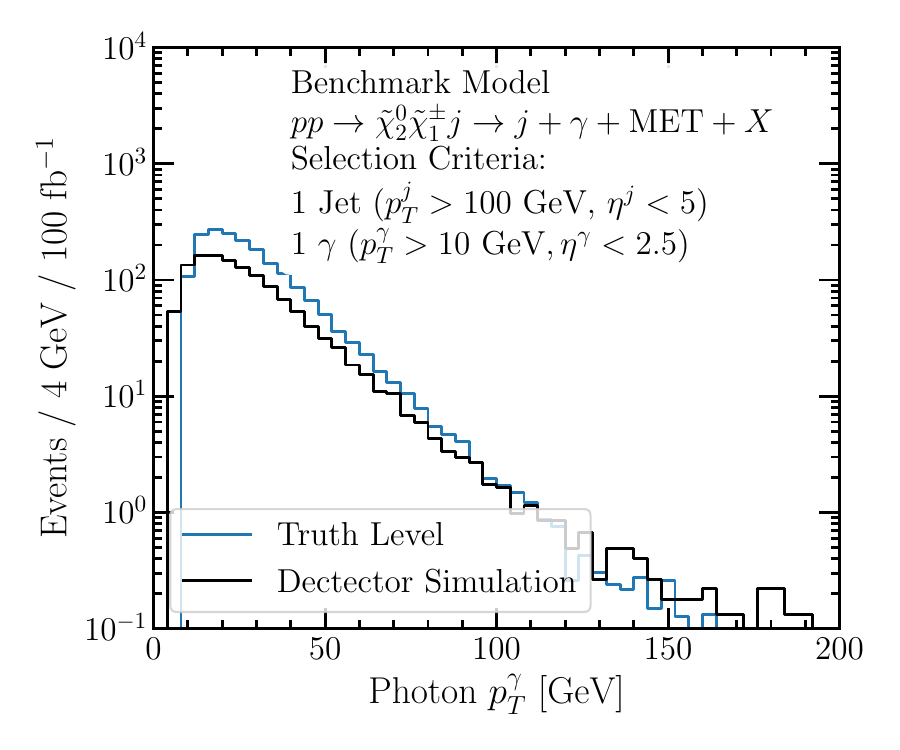}
    \includegraphics[width=0.5\linewidth]{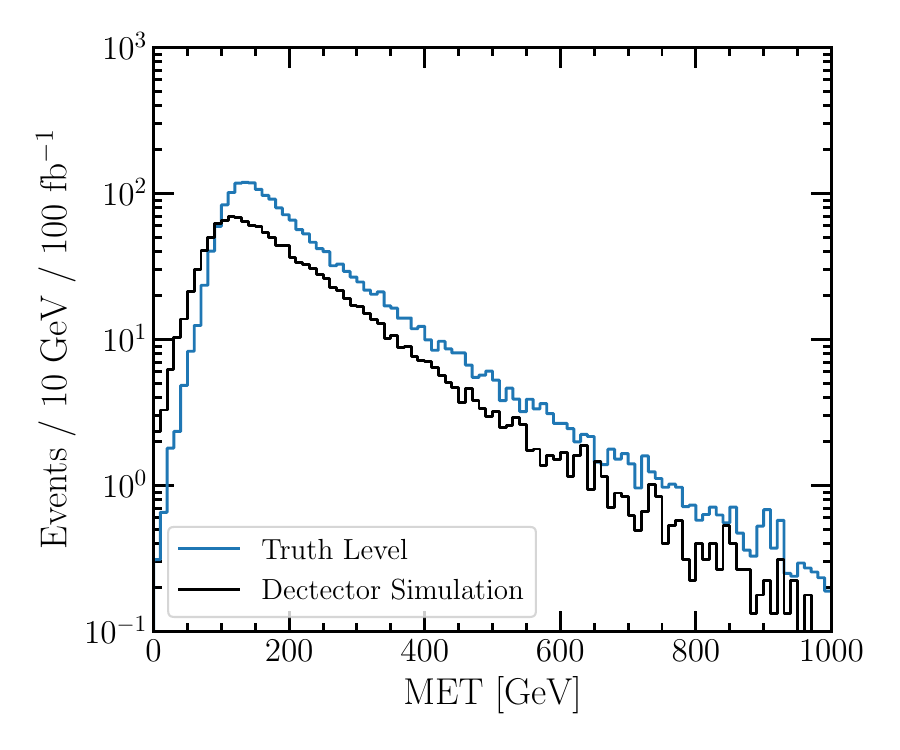}
    \caption{Kinematic distributions of the Benchmark Point generated using a \texttt{Madgraph}+\texttt{Pythia}+\texttt{Delphes} simulation chain. In order to boost the \MET of the event, we consider only ($pp \to \chitwo + \chaone + j$) events with an ISR jet with $p_T^j > 100\,$GeV. We denote the visible decay products of the chargino with ``$X$''; note that while we do not consider the kinematic distribution of these decay products here, they may be useful for a full analysis.}
    \label{fig:simulation}
\end{figure}

To study the kinematic distribution of the objects in the final state, we use the following {\bf benchmark point} featuring a large radiative decay branching ratio of the second-lightest neutralino, a reasonably large ($\massplit$) mass splitting, and providing explanations for DM and $\Delta a_\mu$: 
\begin{center}
\renewcommand{\arraystretch}{1.2}
\begin{tabular}{L{6cm} L{7cm}} 
    \hline
    $M_1 = -282\,$GeV & $m_{\chitwo} \approx m_{\chaone} \approx 300\,$GeV \\
    $M_2 = 286\,$GeV & $m_{\chitwo} - m_{\chione} = 24.1\,$GeV \\
    $\mu = 800\,$GeV & ${\rm BR}(\chitwo \to \chione + \gamma) = 36\,\%$ \\
    $M_{\tilde{l}} = 600\,$GeV & $a_\mu^{\rm MSSM} = 2.0 \times 10^{-9}$ \\
    $\tan\beta = 60$ & $\Omega_{\chione} h^2 = 0.121$ \\
    & $\sigma(pp \to \chaone + \chitwo + j) = 60\,$fb \\
    \hline
\end{tabular}
\end{center}
Note that the production cross section $\sigma(pp\to\chaone+\chitwo+j)$ is the fiducial cross section at the $\sqrt{s} = 13$\,TeV LHC requiring one jet with $p_T^j>100$\,GeV. The NLO contribution to the production cross section is estimated to be 25\%, which is included in this reported cross section, and the kinematic distributions in \fig{fig:simulation} and \fig{fig:kinematics2}. This is motivated by calculations of the $\chargone \chitwo$ production cross sections for this parameter point using \texttt{Prospino 2.1}.

In \fig{fig:simulation}, we show the distribution of \ptgamma and \MET both at the truth level of our simulation chain and after including detector effects via \texttt{Delphes}. Since the ISR jet is the most energetic visible object in the final state, and since the $\chitwo$ and $\chaone$ would be produced with equal and opposite $p_T$ in the absence of initial state radiation, the ($\chitwo$+$\chaone$) system will recoil against the ISR jet in the transverse plane. Since most of the $p_T$ of the second lightest neutralino and the lightest chargino is inherited by the lightest neutralino, the \MET of the event approximately balances the $p_T^j$. Thus, our event selection criterion of $p_T^j > 100\,$GeV (at the level of the hard event) ensures that most of the signal events have $\MET > 100\,$GeV. From \fig{fig:simulation} we can note that our signal events feature a broad high-\MET tail, which allows to select more aggressive \MET cuts in an analysis at moderate cost in selection efficiency. Turning to the \ptgamma distribution shown in the left panel of \fig{fig:simulation}, we can first note that the \ptgamma-distribution does indeed peak just below $(\massplit) \approx 24\,$GeV. The transverse momentum of the photon can however be boosted by an $\mathcal{O}(1)$ factor. This happens if the photon from the ($\chitwo \to \chione + \gamma$) decay is produced in the same direction in which the $\chitwo$ is produced and boosted, leading to the high-$\ptgamma$ tail of the distribution in the left panel of \fig{fig:simulation}.

In order to suppress backgrounds, we anticipate that additional event selection cuts will be necessary. In particular, cuts on $\ptgamma$ more stringent than the $\ptgamma > 10\,$GeV requirement we have made this far may be necessary to sufficiently suppress electromagnetic backgrounds, and additional $\MET$-cuts might be required to suppress backgrounds arising from events with mis-measured jet energies. In \fig{fig:kinematics2}, we present a table of selection efficiencies for signal events if additional cuts on \ptgamma and \MET over our initial criteria ($p_T^j > 100\,$GeV, $p_T^\gamma > 10\,$GeV) are made. For example, requiring $\MET > 150\,$GeV and $\ptgamma > 40\,$GeV leads to a signal event selection efficiency of $16\,\%$. 

\begin{figure}
    \subfloat{
        \raisebox{4.cm}{
            \renewcommand{\arraystretch}{1.1}
            \small
            \begin{tabular}{C{1.1cm} C{0.4cm}| C{0.9cm} C{0.9cm} C{1.1cm}}
                \hline\hline
                & & \multicolumn{3}{c}{\MET cut [GeV]} \\ 
                & & 150 & 300 & 500 \\ \hline 
                \multirow{ 3}{*}{\makecell{$p_T^\gamma$ cut\\ $[$GeV]}} & 0 & 60\% & 17\% & 3.9\% \\ 
                & 40 & 16\% & 6.0\% & 1.8\% \\ 
                & 70 & 3.3\% & 1.7\% & 0.70\% \\ 
                \hline \hline 
            \end{tabular}
            }
        }
    \includegraphics[width=0.6\linewidth]{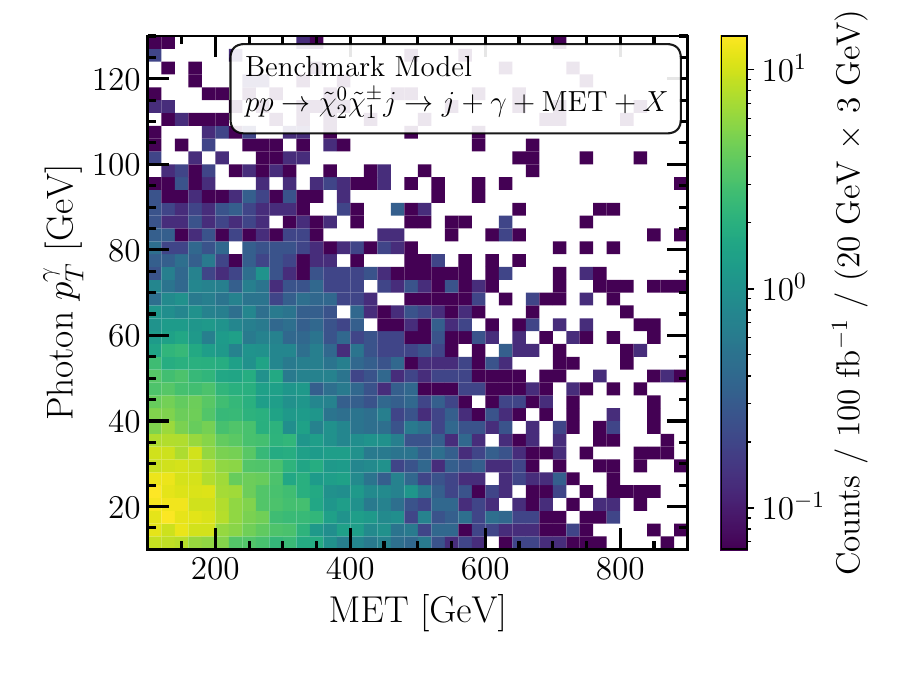}
    \caption{{\it Left:} Efficiency table for various possible \MET and photon $p_T$ cuts applied to our benchmark point. The efficiency is calculated with respect to events that pass the ISR jet and photon requirements listed in bullet points in the main text. {\it Right:} Distribution of photon $\ptgamma$ and \MET at the detector level. There is no apparent correlation between $\ptgamma$ and $\MET$, thus, a multi-object trigger would be well suited for this scenario. Note the logarithmic color scale.}
    \label{fig:kinematics2}
\end{figure}

In the right panel of \fig{fig:kinematics2}, we show a 2D-histogram of the distribution of signal events for our benchmark point in the \MET vs \ptgamma plane. We observe no significant correlation in the distribution of \ptgamma and \MET. Thus, simple cuts on \ptgamma or \MET are likely not the most effective way of searching for such signals. In order to suppress backgrounds sufficiently, we anticipate that a multi-variate analysis of the final state using the kinematics of all visible objects (the photon, the soft visible decay products of the $\chaone$, and the ISR jet) together with \MET will be necessary. Furthermore, the new technical capabilities installed at the LHC during the long shutdown preceding Run~3 now allow for non-trivial multi-object triggers with considerably lower thresholds than more traditional \ptgamma-only or \MET-only triggers. A combined \ptgamma and \MET trigger would be particularly useful to search for the soft-photon + \MET final states arising from $(pp \to \chitwo + \chaone + j)$ events at the LHC when the $\chitwo$ decays radiatively. We note that the analysis presented in this work is aimed at motivating future studies and searches for the yet unexplored radiative decay signatures. To fully utilize these new signatures in collider searches, dedicated studies of the proposed signal and the SM background should be performed to decide the most promising signal topology and the event selection strategies.

Before concluding, let us briefly comment on other interesting collider signature in the region of parameter space discussed in this work. If the strongly charged superpartners, i.e., the squarks and gluinos, are light enough so that their production cross section at the LHC is not too small, then interesting signals could also arise from their production. Wino-like neutralinos can be produced in the decay cascades of squarks and gluinos, and those neutralinos can again decay radiatively, leading to collider signatures at the LHC containing hard jets (from the squark/gluino decays) and photons from radiative decays ($\chitwo \to \chione + \gamma$). In such events, the $\chitwo$'s are produced with much larger boosts  and hence, the photons would have much larger $p_T$ than those arising from the direct $pp \to \chitwo + \chaone (+j)$ production we focus on in this work. We refer the reader to Ref.~\cite{Baer:2005jq} for a discussion of collider signals arising from ($\chitwo \to \chione + \gamma$) decays in the decay cascades of gluinos and squark at the LHC, note that the corresponding final states are comparatively well-covered by exisiting searches. The relevant squark/gluino production cross sections are steep functions of the squark/gluino masses~\cite{Borschensky:2014cia}, and those masses can be adjusted without affecting the phenomenology driven by the electroweak sparticle sector discussed in this work. Hence, while it is paramount that the LHC collaborations continue the search for new particles in the channels aimed at the direct production of strongly charged sparticles, we do not discuss them further in this work.

In summary, the radiative decay of Wino-like neutralinos, ($\chitwo \to \chione + \gamma$), in [$pp \to \chitwo + \chaone (+j)$] events leads to a new potential search channel at the LHC for which we performed a first study in this section: a soft photon accompanied by \MET and additional soft visible decay products arising from $\chaone$ decays. This kinematic region is, to the best of our knowledge, not targeted by any of the current photon + \MET searches at the LHC~\cite{CMS:2018fon, CMS:2017brl, CMS:2017qca,ATLAS:2022ckd, ATLAS:2014kci}. As we have shown throughout this work, a large branching ratio for the ($\chitwo \to \chione + \gamma$) decay is a characteristic feature of an attractive region of the MSSM parameter space. The new search channel proposed here can be especially relevant in complementing the existing multi-lepton searches at the LHC, which are currently most powerful in targeting ($\chitwo + \chaone$) Wino pair-production in the $(\massplit) \sim 10 - 30\,$GeV region. Indeed, the reach of these multi-lepton searches in the region of parameter space we focus on in this work is hampered by the large radiative decay branching ratio. A systematic study of Standard Model backgrounds will be necessary to fully quantify the reach of a soft-photon + \MET search.

\section{Conclusion}
\label{sec:conclusions}

The nature of dark matter is one of the most tantalizing puzzles in theoretical particle physics and calls for new physics beyond the Standard Model. The recent observation of the muon's magnetic moment deviates from current Standard Model expectations and - if such discrepancy persists - also opens an exciting possibility for new physics. In this article, we explore the electroweakino sector of the MSSM which can provide an explanation for dark matter, and in certain cases can simultaneously provide an explanation for the observed value of the muon's magnetic moment. In order to avoid over-closing the universe with the Bino-like dark matter candidate, we focus on the compressed region, where $(\massplit) \sim 10-30\,$GeV, to allow for Bino-Wino co-annihilation. To avoid current constraints from direct dark matter detection experiments, we ask the Higgsino mass parameter $\mu$ to be of opposite sign to the Bino mass parameter $M_1$; this suppresses the spin independent cross section, without the need of a very large $|\mu|$, by cancellations between the different amplitudes contributing to the cross section. The value of $|\mu|$ needs still to be sufficiently large - above a few hundred GeV - to fulfil the bounds from spin dependent direct detection dark matter searches. To obtain a positive contribution to the muon's magnetic moment, on the other hand, we consider the region of parameter space with $\mu$ being of the same sign as the Wino mass parameter $M_2$. Altogether, we concentrate on an attractive region of the MSSM parameter space, with a compressed Bino-Wino spectrum and emphasis on the case of opposite sign gaugino masses due to $(\mu \times M_1) < 0$ and $(\mu \times M_2) > 0$. This region not only provides a viable DM candidate and may explain the measured value of the magnetic dipole moment of the muon, but is also in accordance with current dark matter direct detection and collider constraints.

To explore a new discovery opportunity for Bino-like dark matter at the LHC, we first revisit the collider phenomenology in the compressed region, and show that for $(M_1 \times M_2) < 0$, there is an enhancement of the rate of the radiative decay of the second-lightest neutralino into the lightest neutralino and a photon. This signal emerges as a frequent and yet unexplored physical process. The enhanced radiative decay rate leads to a suppression of the decays of the second-lightest neutralino into charged leptons, weakening the reach of the LHC multi-lepton searches. In this opposite-sign gaugino mass scenario, the region with $200\,{\rm GeV}\lesssim m_{\tilde{\chi}_2^0} \lesssim 350\,$GeV remains unconstrained by the existing experimental searches, while most of the parameter space in the same-sign gaugino mass scenario, $(M_1 \times M_2) > 0$, has been excluded. 

We propose a new channel to search for ($pp \to \chitwo + \chaone$) production in the Wino-Bino compressed region at the LHC: mono-photon+\MET accompanied by jets or a charged lepton from the chargino decay, and possibly an initial state radiation jet. We have given estimates of the signal cross sections, kinematic distributions, and selection efficiencies relevant for such a search, indicating that this channel could complement the current leptonic searches at the LHC. A dedicated study of this potential signal should be performed by the ATLAS/CMS collaborations. Ultimately, a search for radiative decays of the second-lightest neutralino could allow a deeper exploration of minimal supersymmetric models that provide explanations for both dark matter and the muon's magnetic moment anomaly, and perhaps, lead to the discovery of new particles at the LHC.

\section{Acknowledgements} 
We would like to thank David Miller for useful discussions and comments. 
SB is supported in part by NSF Grant PHY-2014215, DOE HEP QuantISED award \#100495, and the Gordon and Betty Moore Foundation Grant GBMF7946. 
Fermilab is operated by Fermi Research Alliance, LLC under Contract No. DE-AC02-07CH11359 with the U.S. Department of Energy. 
The work of CW at the University of Chicago has been also supported by the DOE grant DE-SC0013642. 
Work at ANL is supported in part by the U.S. Department of Energy (DOE), Div. of HEP, Contract DE-AC02-06CH11357. 
This work was supported in part by the DOE under Task TeV of contract DE-FGO2-96-ER40956. 
NRS is supported by U.S. Department of Energy under Contract No. DE-SC0007983. 
This work was performed in part at Aspen Center for Physics, which is supported by National Science Foundation grant PHY-1607611. TO is supported by the Visiting Scholars Program of URA.

\newpage
\begin{appendices}\label{app:Appendices}

\section{Analytic Calculations of Radiative Decay}
\label{app:radiative_decay}

The loop integrals used for the analytic calculation of the neutralino radiative decay in \sect{sec:decay} are given as following:
\begin{eqnarray}
        I_2 &=& \frac{1}{\Delta}\int_0^1 dx \log X \;, \\
        I &=&\frac{1}{\Delta}\int_0^1\frac{dx}{1-x}\log X \;, \\
        J &=& I(X\to X') \;, \\
        K &=&-\frac{1}{\Delta}\int_0^1 dx\left(1+\frac{B}{\Delta x(1-x)}\log X\right) \,.
    \end{eqnarray}
    The parameters $\Delta$, $B$ and $X (X')$ are defined as
    \begin{eqnarray}
        \Delta &\equiv& m_{\chitwo}^2-m_{\chione}^2 \;, \\
        B &\equiv& x m_{f}^2 +(1-x) m_{b}^2-x (1- x)m_{\chitwo}^2 \;, \\
        X&\equiv&\frac{x m_{f}^2 +(1-x)m_{b}^2-x(1-x)m_{\chitwo}^2}{x m_{f}^2 +(1-x)m_{b}^2-x (1-x)m_{\chione}^2} \;, \\
        X' &\equiv& X(m_b\leftrightarrow m_{f}) \;,
    \end{eqnarray}
    where $m_b$ and $m_f$ are the masses of the boson and fermion in the loop, respectively.

\section{Further Plots of $m_{\chitwo}$ vs. $\Delta m(\chitwo, \chione)$ 
\label{app:Appendix}}

\begin{figure}[H]
    \vspace{-1cm}
    \centering
    \includegraphics[width=\textwidth]{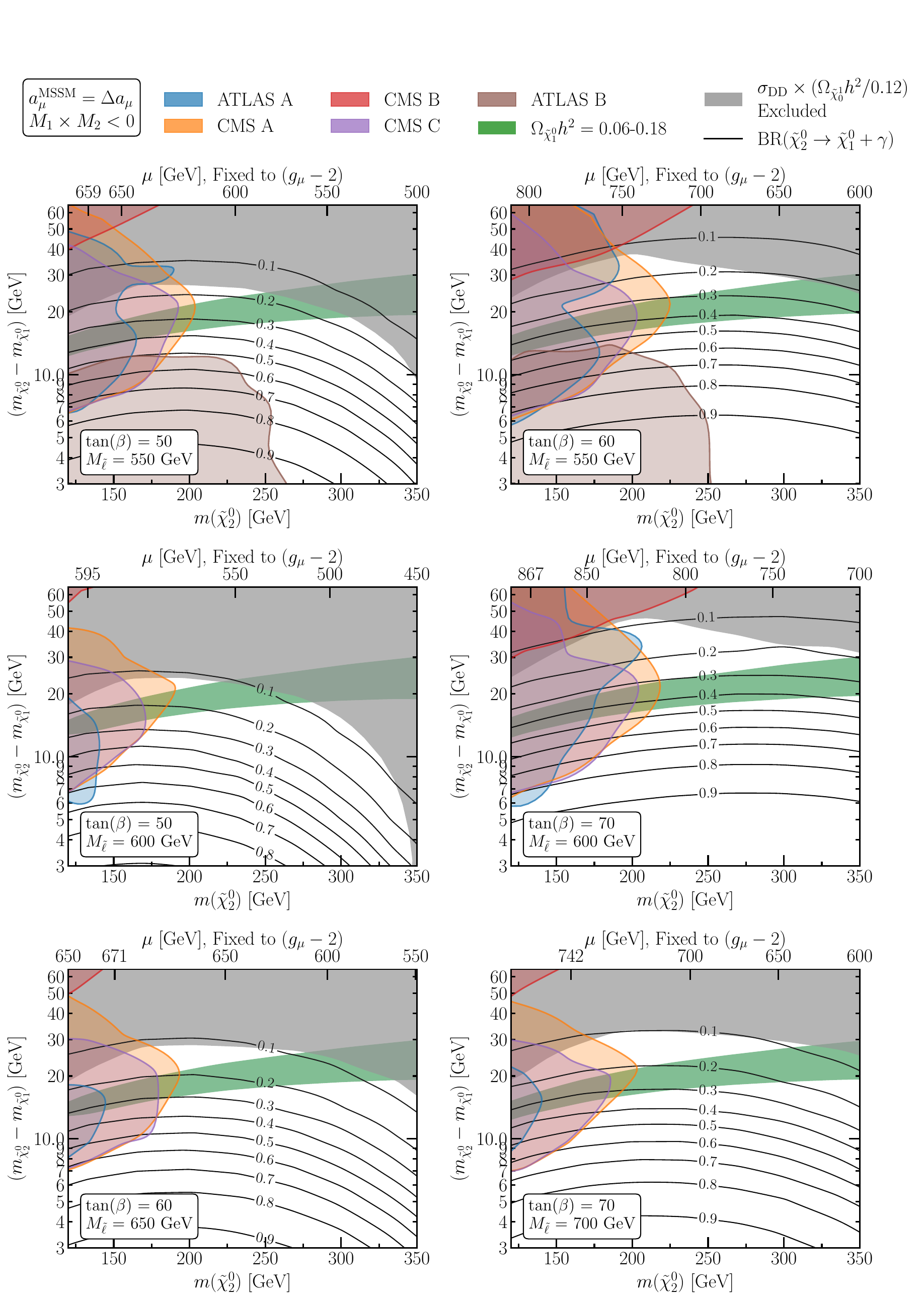}
    \caption{Same as Fig.~\ref{fig:exclusion_60_600}, but for different choices of soft slepton mass parameter $M_\slepton$ and $\tan(\beta)$, as denoted in each panel. Note that all panels are for $(M_1 \times M_2) < 0$, and throughout we adjust $\mu$ such that $\Delta a_\mu = a_\mu^{\rm MSSM}$.}
    \label{fig:more_plots}
\end{figure}

In \fig{fig:more_plots}, we show the parameter space satisfying $\Delta a_\mu = a_\mu^{\rm MSSM}$, for various choices of slepton masses and $\tan\beta$. Here, we focus on the case of $(M_1\times M_2)<0$, for which direct detection constraints are alleviated and the radiative decay branching ratio of the second-lightest neutralino is enhanced. \fig{fig:more_plots} indicates that direct detection constraints are weaker for larger $\tan\beta$ and for smaller slepton masses, mainly because then larger $\mu$ is needed to explain $\Delta a_\mu$. Note that for a lighter slepton mass, $M_{\slepton}=550$\,GeV, the search for sleptons and charginos production \cite{ATLAS:1908.08215} (``ATLAS B" in the figure) rules out the region with small $m_\chitwo$ and $(\massplit)$, as expected from the increased slepton production cross section due to its reduced mass.

\vspace{2cm}

\end{appendices}
\bibliography{ref}

\end{document}